\documentclass{pasj01}
\renewcommand{\rm}[1]{\mathrm{#1}}
\renewcommand{\ss}[1]{${}^{#1}$}
\newcommand{\cd}[3]{$#1(#2)\times10^{#3}$}

\draft
\Received{$\langle$reception date$\rangle$}
\Accepted{$\langle$acception date$\rangle$}
\Published{$\langle$publication date$\rangle$}

\begin{document}
\title{An unbiased spectral line survey observation toward the low-mass star-forming region L1527}
\author{Kento \textsc{Yoshida},\altaffilmark{1,2,}$^{*}$ Nami \textsc{Sakai},\altaffilmark{2} Yuri \textsc{Nishimura},\altaffilmark{1,3,4} Tomoya \textsc{Tokudome},\altaffilmark{1} Yoshimasa \textsc{Watanabe},\altaffilmark{5,6} Takeshi \textsc{Sakai},\altaffilmark{7} Shuro \textsc{Takano},\altaffilmark{8} and Satoshi \textsc{Yamamoto}\altaffilmark{1,9}}

\altaffiltext{1}{Department of Physics, The University of Tokyo, 7-3-1, Hongo, Bunkyo-ku, Tokyo 113-0033, Japan}
\altaffiltext{2}{Star and Planet Formation Laboratory, RIKEN Cluster for Pioneering Research (CPR), 2-1, Hirosawa, Wako, Saitama 351-0198, Japan}
\altaffiltext{3}{Institute of Astronomy, The University of Tokyo, 2-21-1, Osawa, Mitaka, Tokyo 181-0015, Japan}
\altaffiltext{4}{Chile Observatory, National Astronomical Observatory of Japan, 2-21-1, Osawa, Mitaka, Tokyo 181-8588, Japan}
\altaffiltext{5}{Division of Physics, Faculty of Pure and Applied Sciences, University of Tsukuba, Tsukuba, Ibaraki 305-8571, Japan}
\altaffiltext{6}{Tomonaga Center for the History of the Universe, University of Tsukuba, Tsukuba, Ibaraki 305-8571, Japan}
\altaffiltext{7}{Department of Communication Engineering and Informatics, Graduate School of Informatics and Engineering, The
University of Electro-Communications, Chofugaoka, Chofu, Tokyo 182-8585, Japan}
\altaffiltext{8}{Department of Physics, General Studies, College of Engineering, Nihon University, 1, Nakagawara, Tokusada, Tamuramachi, Koriyama, Fukushima 963-8642, Japan}
\altaffiltext{9}{Research Center for the Early Universe, The University of Tokyo, 7-3-1, Hongo, Bunkyo-ku, Tokyo 113-0033, Japan}
\email{yoshida@taurus.phys.s.u-tokyo.ac.jp}

\KeyWords{astrochemistry --- ISM: individual (LDN 1527) --- ISM: molecules}

\maketitle
\begin{abstract}
An unbiased spectral line survey toward a solar-type Class 0/I protostar, IRAS04368+2557, in L1527 has been carried out in the 3 mm band with the Nobeyama 45 m telescope. L1527 is known as a warm carbon-chain chemistry (WCCC) source, which harbors abundant unsaturated organic species such as $C_n H\ (n=3,\ 4,\ 5,\ldots)$ in a warm and dense region near the protostar. The observation covers the frequency range from 80 to 116 GHz. A supplementary observation has also been conducted in the 70 GHz band to observe fundamental transitions of deuterated species. In total, 69 molecular species are identified, among which 27 species are carbon-chain species and their isomers, including their minor isotopologues.  This spectral line survey provides us with a good template of the chemical composition of the WCCC source.
\end{abstract}

\section{Introduction}
In the last decade, it has been established that the chemical composition of low-mass protostellar sources shows significant diversity, even if their evolutionary stages are similar to one another (e.g., \cite{sy13}; \cite{watanabe12}). One distinct case is hot corino chemistry characterized by rich saturated complex organic molecules (COMs) in a hot region around a protostar (e.g., \cite{cazaux03}; \cite{s06}; \cite{oberg11}; \cite{taquet15}; \cite{codella16}). A representative hot corino source is IRAS 16293-2422 in Ophiuchus (e.g., \cite{cazaux03}; \cite{bottinelli04}; \cite{kuan04}). Recent ALMA observations revealed that saturated COMs such as HCOOCH$_3$, (CH$_3$)$_2$O, and even glycolaldehyde are abundant in the innermost part of the protostellar core having a temperature higher than 100 K (e.g., J{\o}rgensen et al. \yearcite{jorgensen12,jorgensen16}; \cite{favre14}; Oya et al. \yearcite{oya16, oya18}). On the other hand, the warm carbon-chain chemistry (WCCC) is characterized by high abundances of carbon-chain molecules and related species concentrated around a protostar (e.g., Sakai et al.\ \yearcite{s08b,s09a,s10a}; \cite{hirota10}). A prototypical source is IRAS 04368+2557 in L1527 in the Taurus molecular cloud. In WCCC, unsaturated hydrocarbons such as carbon-chain molecules and their isomers are efficiently produced in the gas phase in a lukewarm region around a protostar ($T\sim30$ K and $R\sim1000$ au).  Their production is triggered by sublimation of CH$_4$ from grain mantles.

Hot corino chemistry and WCCC show exclusive nature: carbon-chain molecules are deficient in the hot corinos, while COMs are deficient in the WCCC sources \citep{s08b,sy13,lefloch18}.
It is proposed that this chemical variation would originate from the duration time of starless core phase after shielding of the interstellar UV radiation from outside of the parent molecular cloud \citep{s09a, sy13}.
If the duration time is long enough for the gas-phase formation of CO from carbon atom, CO is depleted onto dust grains rather than C atom, resulting in the formation of CH$_3$OH through a series of reactions with the H atom on dust surface (e.g., \cite{wk02};  \cite{soma15}). COMs are thought to be formed on dust surfaces as in the case of CH$_3$OH, or by subsequent gas-phase reactions starting from sublimated CH$_3$OH (e.g., \cite{gh06}; \cite{vh13}; \cite{balucani15}; \cite{soma18}). On the other hand, if the duration time of the starless core phase is close to the free-fall time (i.e., the core collapse starts just after the UV shielding), the C atom still survives in the gas phase without being converted to CO, and can be depleted onto dust grains. This situation leads to the efficient production of CH$_4$ on dust surfaces by hydrogenation of the C atom \citep{aikawa08}. After the onset of star formation, molecules in dust mantles are released into the gas phase, and chemical diversity emerges in the gas phase.

A detailed understanding of the origin and the fate of the chemical diversity is of great interest for astrochemistry in relation to the origin of the Solar System. As the first step toward this goal, it is important to reveal the chemical compositions of representative low-mass protostellar sources in an unbiased way. Although spectral line survey observations toward the hot corino source IRAS 16293-2422 have been conducted not only with single-dish telescopes but also with interferometers \citep{caux11, jorgensen16, ligterink17}, no spectral line survey observations toward WCCC sources have been reported except for that recently published by \citet{lefloch18} (See below).
With this in mind, we conducted the unbiased spectral line survey toward the prototypical WCCC source L1527 in the 3 mm band with the Nobeyama 45m telescope (hereafter referred to as NRO 45 m), as part of the legacy project of Nobeyama Radio Observatory. Here, we report the whole result of the survey. Presence of highly unsaturated hydrocarbons is the most characteristic feature of interstellar chemistry. Hence, revealing the whole chemical composition of the WCCC source and its comparison with other sources such as starless cores showing cold carbon-chain chemistry (e.g., TMC-1 (CP); \cite{kaifu04}) will give us valuable information on formation of these molecules in space.

As an independent study, a similar spectral line survey observation toward this source was also carried out with the IRAM 30 m telescope, as part of the ASAI (Astrochemical Surveys At Iram) program. Its summary result has recently been reported \citep{lefloch18}, although detailed data exploitation has not been presented. The ASAI survey almost covers the frequency range of our survey in the 3 mm band with a larger beam size by a factor of 1.5. Nevertheless, independent spectral line surveys with different telescopes are always important for the fundamental sources such as L1527, because they give opportunities to confirm the consistency of the results, and also provide additional information such as the source sizes by taking advantage of the different beam sizes. Here, we report the results of our spectral line survey with NRO 45 m.

\section{Observation}
The spectral line survey observations of L1527 were carried out with NRO 45 m during the seasons from 2006 to 2012.
The frequency range was from 79.8 GHz to 116.8 GHz. The observed position was $(\alpha_{2000},\  \delta_{2000})=(\timeform{4h39m53.89s},\ \timeform{26D03'11.0''})$, which is the position of the protostar \citep{s08b}. In this observation, we used the SIS mixer receivers, S80 and S100, simultaneously from 2006 to 2008, while we employed the dual-polarization sideband-separating SIS receiver T100H/V \citep{nakajima08} from 2009 to 2012. The system temperatures varied from 250 to 350 K in the former and from 150 to 250 K in the latter. The intensity scale was calibrated by using the chopper-wheel method, and the calibration uncertainty is estimated to be better than 20\%. The telescope pointing was checked once every 1 to 1.5 hour by observing the nearby SiO maser sources (NML-Tau and Ori-KL). The pointing accuracy was ensured to be better than $6''$. The main-beam efficiency ($\eta_\mathrm{mb}$) at 86 GHz was 0.43 from 2006 to 2008, 0.49 in 2009, and 0.42 from 2010 to 2012, as reported on the NRO website. The antenna temperature ($T_\mathrm{A}^*$) is converted to the main-beam brightness temperature ($T_\mathrm{mb}$) by $T_\mathrm{mb}=T_\mathrm{A}^*/\eta_\mathrm{mb}$. The beam size was $19''$ at 86 GHz and $15''$ at 110 GHz. A position switching mode with the off-position $(\alpha_{2000},\  \delta_{2000})=(\timeform{4h42m35.9s},\ \timeform{25D53'23.3''})$, which is free from the emission of CO isotopologue lines, was employed in all the observations.

Before 2010, we used a bank of acousto-optical radio spectrometers (AOSs), whose bandwidth, resolution, and channel spacing are 250 MHz, 250 kHz, and 125 kHz, respectively. In 2011-2012, we used a bank of autocorrelators, SAM45 (Spectral Analysis Machine for the 45 m telescope), whose bandwidth, resolution, and channel spacing were set to be 1 GHz, 244 kHz, and 244 kHz, respectively. The frequency resolution corresponds to the velocity resolution of $\sim0.8$ km s$^{-1}$ at 90 GHz. This resolution is larger than the typical line width in this source ($\sim0.5$--$0.8$ km s$^{-1}$). Nevertheless, we chose this resolution to cover the whole 3 mm band within the limited observation time. Although the lines are partly frequency-diluted, the integrated intensity is reliable.

In addition to the observations in the 3 mm band, a supplementary observation was conducted in the 4 mm band with NRO 45 m from February to April 2012. This observation mainly aims at investigating the deuterium fractionation of some fundamental molecular species. Thus, only selected lines were observed. The side-band-separating (2SB) SIS mixer receiver T70H/V was used as the front end with a typical system noise temperature of 200--300 K. The back end was SAM45, whose resolution was set to be 60.1 kHz. The beam size was $22''$ at 75 GHz. The main-beam efficiency at 75 GHz was 0.45.
\section{Results}
\subsection{Overall results}
\begin{figure}[t]
\centering
\includegraphics[bb=0 0 2400 1200, width=.98\textheight, angle=90]{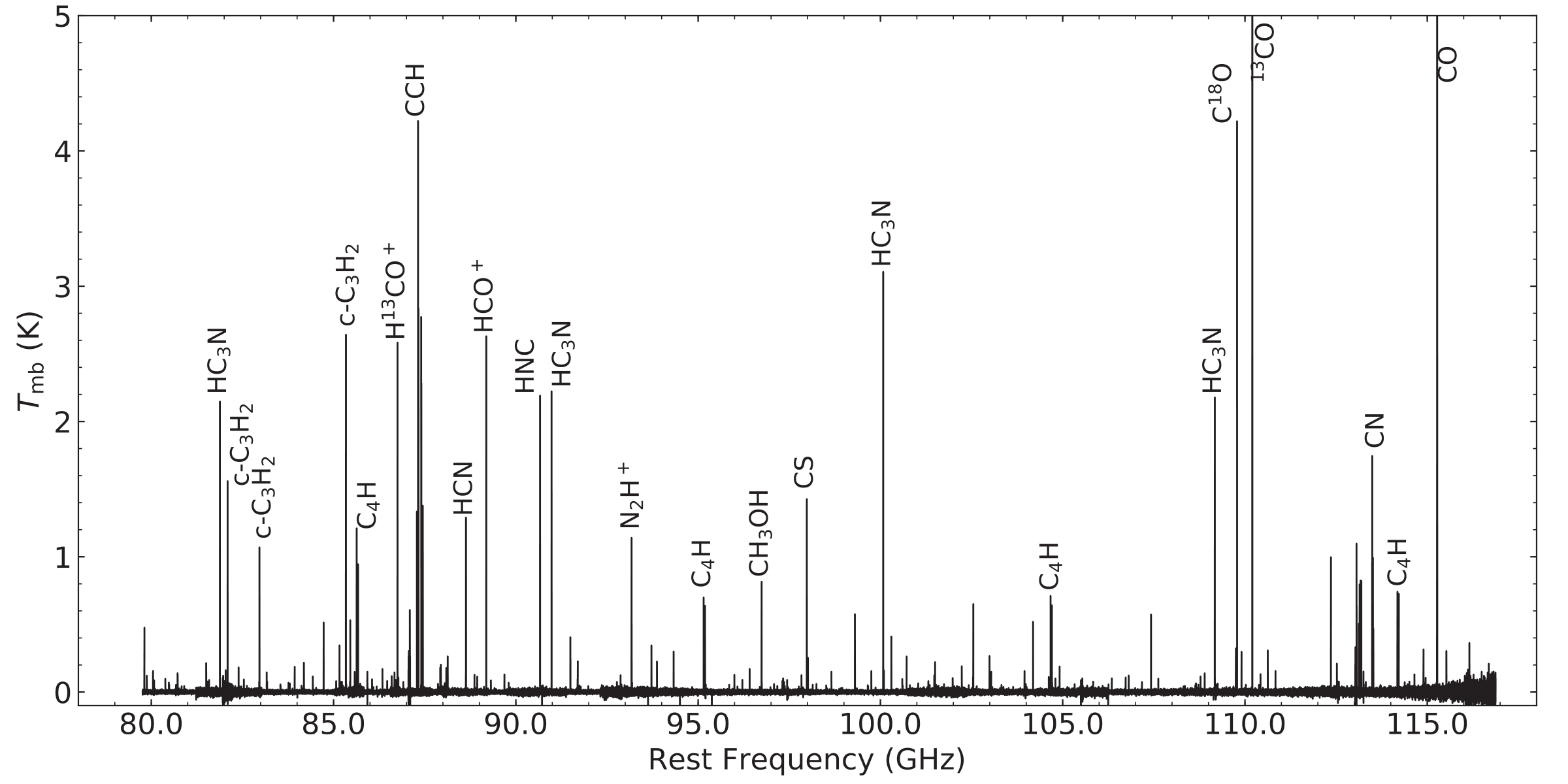}
\caption{The overall spectrum of L1527 in the 3 mm band. Lines of some representative molecular species are indicated.}\label{spect_3mm_ov}
\end{figure}

Figure \ref{spect_3mm_ov} is an overview of the observed spectrum in the 3 mm band, while Figure \ref{spect_3mm} shows its zoomed view. Figure \ref{spect_4mm} shows the spectra of the supplementary observation in the 70 GHz band.  Typically, the rms noise at the native spectral resolution ranges from 5 to 15 mK in $T_\mathrm{mb}$ in the 3 mm band. This rms noise level is similar to or even better than that of the spectral line survey toward the outflow shocked region L1157 B1, which was also conducted with NRO 45 m as part of the legacy project of Nobeyama Radio Observatory (\cite{sugimura11}; \cite{yamaguchi12}).
The detected lines are identified on the basis of spectral line databases, the Cologne Database for Molecular Spectroscopy managed by University of Cologne (CDMS; \cite{muller01}; \yearcite{muller05}) and the Submillimeter, Millimeter, and Microwave Spectral Line Catalog provided by Jet Propulsion Laboratory (JPL; \cite{pickett98}). A line detection criterion is that the peak intensity of the line exceeds four times the rms noise level at its expected frequency. In total, 243 emission lines and one absorption line (CH$_3$OH$\;3_{1,3}$--$4_{1,4},\;$A$^+$ at 107.01 GHz) are detected in the frequency range from 79.8 to 116.9 GHz (Figures \ref{spect_3mm_ov} and \ref{spect_3mm}). Hence, the line density is 6.5 GHz$^{-1}$ with this sensitivity. From the detected emission lines, 69 molecular species are identified in the 3 and 4 mm bands, among which 37 species are isotopologues. These numbers are higher than those of L1157 B1 (line density: 3.4 GHz$^{-1}$, detected species: 47 species including 15 isotopologues).  This clearly shows chemical complexity of L1527. The detected molecules are summarized in Table \ref{tab_detected}.

In this survey, c-C$_3$D is detected for the first time in interstellar clouds.  Although c-C$_3$D is detected with the signal-to-noise (S/N) ratio higher than $4\sigma$, the velocity resolution is not high enough to determine the line parameters by Gaussian fits. C$_2$O, C$_3$S, C$_6$H, HCNO, HCO$_2^+$, C$_4$D, C$_2$H$_3$CN, and CH$_2$DOH are tentatively detected with the S/N ratio higher than $3\sigma$. In addition, CC\ss{13}CCH, l-C$_3$D and CH$_3$CCD are also tentatively detected, because multiple lines are marginally seen.  Among the tentatively detected species in this survey, the detections of C$_6$H, HCNO, HCO$_2^+$, C$_4$D, and l-C$_3$D have already been reported by the higher sensitivity observations \citep{s07, marcelino09, s08c, s09b}.

For unidentified lines with the S/N ratio higher than 4$\sigma$, we carefully inspect whether they are also detected in the ASAI survey, and find that none of them have corresponding features in the ASAI data. Most of the unidentified lines that appeared only in this survey are likely spurious lines, which are mainly caused in the AD converters of the autocorrelators during the period of our observations. Significant features of such spurious lines are shown in some panels of Figure \ref{spect_3mm}. 
In the frequency range observed only with NRO 45 m, 2 lines are unidentified, and they should be verified in future observations. The intensity, line-of-sight velocity and FWHM line width of the detected lines are determined by a single Gaussian fit. If the line profile is blended with multiple lines such as nearby hyperfine components, multiple Gaussian functions are employed to determine the line parameters. For the hyperfine components of N$_2$D$^+$ and CH$_2$CN, we assume that the line width and the line-of-sight velocity are identical among all the components, and that the intensities are proportional to the line strengths, because the hyperfine splittings are too small to fit them independently (Figure \ref{N2D_CH2CN}). Tables \ref{line_parameter_3mm} and \ref{line_parameter_4mm} present lists of the line parameters of the identified molecular lines including tentatively detected ones and unidentified lines.  Note that some of the hyperfine components of CCD show significant offsets in $\Delta V_\mathrm{LSR}$ (Table \ref{line_parameter_4mm}). This is probably because of insufficient accuracy of the rest frequencies listed in CDMS (See the Appendix).

As listed in Table \ref{tab_detected}, 27 species are carbon-chain species and their isomers. For saturated molecules, the most fundamental species, CH$_3$OH and CH$_3$CHO, which are already detected in many cold starless cores (e.g., \cite{kaifu04}; \cite{vastel14}; \cite{soma18}), are detected. However, larger complex organic molecules such as (CH$_3$)$_2$O and HCOOCH$_3$ are not detected in this survey, while spectral lines of these species are detected in the 3 mm band with peak intensities of a few 10 mK or stronger toward the hot corino sources such as IRAS 16293-2422 and NGC1333 IRAS4A (e.g., \cite{caux11}; \cite{bottinelli04}). A deficiency of complex organic molecules in L1527 was also reported on the basis of high sensitivity observations \citep{s08a}.

\subsection{Column densities and rotation temperatures}
In this observation, multiple transition lines with different upper-state energies were detected for many species. Assuming a local thermodynamic equilibrium (LTE) condition, we determine the rotation temperature and the beam-averaged column density for each species by using the least-squares method with the following formula:
\[T_\mathrm{b} = \frac{h\nu}{k}\left[\frac{1}{\exp(h\nu/kT_\mathrm{rot})-1}-\frac{1}{\exp(h\nu/kT_\mathrm{bg})-1}\right](1-e^{-\tau}),\]
and
\[\tau=\frac{8\pi^3S\mu^2}{3h\Delta v U(T_\mathrm{rot})}\left[\exp\left(\frac{h\nu}{kT_\mathrm{rot}}\right)-1\right]\exp\left( -\frac{E_\mathrm{u}}{kT_\mathrm{rot}} \right)N,\]
where $T_\mathrm{b}$ is the brightness temperature, $h$ the Planck constant,  $\nu$ the transition frequency, $k$ the Boltzmann constant, $T_\mathrm{rot}$ the rotation temperature, $T_\mathrm{bg}$ the cosmic microwave background temperature of 2.7 K, $\tau$ the optical depth,  $S$ the line strength, $\mu$ the dipole moment, $\Delta v$ the line width at half maximum, $U(T_\mathrm{rot})$ the partition function, $E_\mathrm{u}$ the upper state energy, and $N$ the total column density.

The partition function is numerically calculated from energies and degeneracies of the rotational levels, which are taken from the spectroscopy databases CDMS and JPL.
The beam-filling factor is not included for simplicity. This assumption is reasonable especially for carbon-chain molecules, because distributions of carbon-chain molecules such as CCH and C$_4$H are reported to be extended over $20''$--$40''$ \citep{s10a}, which is comparable to or larger than the typical beam size of the observation ($\sim20''$).
Uncertainties of the derived column densities include the rms noise, the fitting error, and the intensity calibration uncertainty of 20\%.
Table \ref{tab_NT} shows the column densities and the rotation temperatures derived by the above analysis.
The large uncertainty for C$_3$N originates from large scattering of the data due to the poor S/N ratios.

For molecules for which only one transition line or multiple lines with almost the same upper-state energies are detected, the column densities are derived under the LTE assumption with the excitation temperatures of 10 K and 15 K. The range of the excitation temperature is chosen on the basis of the rotation temperatures derived for other molecules in the above analysis. The column densities thus evaluated are summarized in Table \ref{tab_N_fix}. A change in the temperature by 5 K does not result in a substantial change in the column density ($\sim20$\% or less for most of the species). Optical depths of the lines are also listed in Table 5.  Most of them are less than 1.  Hence, the column densities are reasonably estimated. An exception is the HCCNC case. The column density of HCCNC changes by a factor of 2 between 10 K and 15 K. This is likely due to the relatively high upper-state energy of the detected line ($J=10$--$9;\  E_\mathrm{u}=26$ K). Moreover, if we employ the $J=9$--8 line  ($E_\mathrm{u}=21$ K) tentatively detected with the $3\sigma$ confidential level, the column density is derived to be about a half of that from  the $J=10$--9 line, assuming the excitation temperature of 10 K or 15 K. The column density of HCCNC therefore has a relatively large uncertainty.

It should be noted that some molecules having an extended distribution might have the excitation temperature less than 10 K, because of contributions from less dense and colder parts. However, the excitation temperature of 5 K does not cause a substantial change in the column density ($\sim20$\% or less) and the optical depth (0.4 or less) for most of the species. For DNC and H\ss{13}CO$^+$, the column densities becomes $\sim2$ times higher and optical depths becomes $\sim2$, if the excitation temperature were 5 K.

The low excitation temperature indicates that the emissions are not truly in the LTE condition, but are sub-thermally excited.
To examine whether LTE is a good approximation for this observation, we compare the derived column densities with those obtained with the non-LTE radiative transfer code RADEX \citep{vdT07}. Simple molecules such as C\ss{18}O, CS, SO, HC\ss{15}N, HC\ss{18}O$^+$, and CCH are examined, for which the collisional cross sections are available. As summarized in Table \ref{tab_NLTE}, the column densities derived by RADEX and those derived by the LTE analysis agree within the error over a range of $n(\mathrm{H_2})=3\times10^5$--$3\times10^6\ \mathrm{cm^{-3}}$ and $T_\mathrm{k}=10$--$30$ K except for CS.
The difference in the column density of CS is about a factor of 2, which does not change the conclusion of the discussion in Section 4.2 (see Figures \ref{N_tmc1} and \ref{N_16293}).
\citet{y15} reported the column density of c-C$_3$H$_2$ with RADEX by using the NRO 45 m and the IRAM 30 m telescope, which only differ from those derived with LTE by $20\%$.
Larger molecules tend to exist in the denser part of the cloud, and hence, the non-LTE effect could be smaller than the above cases. The non-LTE effect on isotopic ratios is also negligible, because uncertainties of the derived column densities are mostly caused by assumptions of temperature and H$_2$ density, and their effects are canceled out in the ratios. Thus, we employ the results obtained with LTE in the following analysis and discussions.

The spectral lines of CO, ${}^{13}$CO, CS, HCO$^+$, HCN, and HNC are optically thick, judging from the isotopologue lines. Hence, the column densities of these molecules shown in Table \ref{tab_N_fix} are derived from the isotopologue lines, assuming the following isotopic ratios in the local interstellar matter (ISM): \ss{12}C/\ss{13}C$=60$, \ss{16}O/\ss{18}O$=560$, and \ss{32}S/\ss{34}S$=22$ \citep{LL98, WR94, chin96}.
As discussed later (Section \ref{sect13C}), the \ss{12}C/\ss{13}C ratios of carbon-chain molecules are higher than the local ISM value. However, we here employ the standard values for the above fundamental species.
Although the optical depths of C\ss{18}O and H\ss{13}CO$^+$ lines are also high, the derived column densities are consistent with those derived from the  C\ss{17}O, \ss{13}C\ss{18}O, and HC\ss{18}O$^+$ lines (\ss{18}O/\ss{17}O$=3.2$; \cite{WR94}).

For molecules having the ortho and para species due to symmetry, they are analyzed separately under the assumption of the same rotation temperatures, if the lines of both species are detected. Free parameters for the fit are the total column density, the rotation temperature, and the ortho-to-para ratio. For l-C$_3$H$_2$, c-C$_3$H$_2$, C$_4$H$_2$ and CH$_2$CO, the ortho-to-para ratios are derived to be $2.81\pm0.14$, $2.2\pm0.3$, $3.2\pm0.2$ and $3.2\pm1.2$, respectively. c-C$_3$H$_2$ shows lower ortho-to-para ratio than 3, probably because an optically-thick ortho line is used for the evaluation.
For molecules of which only ortho lines are detected, we assume the statistical ortho-to-para ratios: 3 for H$_2$\ss{13}CO, H$_2$CS, NH$_2$D, c-H$_2$C$_3$O, c-\ss{13}CCCH$_2$ and CH$_2$CN, and 2 for D$_2$CO and c-C$_3$D$_2$. For CH$_3$CCH and CH$_3$CN, the $A$ and $E$ states caused by the internal rotation are analyzed separately under the assumption of the identical rotation temperature. The $A/E$ ratios are derived to be $0.98\pm0.14$ and $1.07\pm0.02$ for CH$_3$CCH and CH$_3$CN, respectively. For CH$_3$CHO and CH$_3$OH, the $A/E$ ratio is assumed to be 1, because only one line of the $A$ state is detected.  The column densities shown in Tables \ref{tab_NT} and \ref{tab_N_fix} represent the total column densities. The error of the column density of NH$_2$D is large because of the poor S/N ratio of the detected line.

We derive the upper limit to the column densities for important undetected species, OCS, HCOOCH$_3$, and (CH$_3$)$_2$O, to compare them with those derived in IRAS 16293-2422.
A spectral line for which the highest S/N ratio is expected in the observed frequency range (i.e., $J=7$--6 at 85.14 GHz for OCS, $4_{14}$--$3_{13}\,\rm{EE}$ at 99.32 GHz for (CH$_3$)$_2$O, and $9_{09}$--$8_{08}\,\rm{A}$ at 100.68 GHz for HCOOCH$_3$) is used for the evaluation for each species. The 3$\sigma$ upper limit is derived from the rms noise under the assumption of the line width of 1.0 km s$^{-1}$, as shown in Table \ref{tab_N_fix}.

In order to derive the beam-averaged fractional abundances of molecules relative to H$_2$ ($X = N/N(\mathrm{H_2})$), we evaluate the beam-averaged column density of H$_2$, $N(\rm{H_2})$, from the C\ss{17}O data. The $N(\rm{H_2})$ value is derived to be $(8.2\pm1.6)\times10^{22}$ cm$^{-2}$ by assuming the rotation temperature of 15 K and the $N(\mathrm
{C{}^{17}O})/N(\mathrm{H_2})$ ratio of $2.6\times10^{-8}$ \citep{jorgensen02}. Note that this ratio is derived from the C\ss{17}O and dust continuum observations, and is lower than the ratio reported by \citet{frerking82}.
The fractional abundances of molecules are listed in Table \ref{tab_X}. Here, we adopt the column densities derived with the excitation temperature of 15 K for the species listed in Table \ref{tab_N_fix}.
\section{Discussion}
\subsection{Isotopic ratios}
\subsubsection{Deuterated species}
Since we detect 12 deuterated species with reasonably good S/N ratios, we here derive their abundance ratio relative to the normal species, as summarized in Table \ref{Dratio}. For the isotopic species, we adopt the same rotation temperatures as those for the normal species. For molecules for which the rotation temperatures are not derived by the fit, the ratios are derived by assuming the excitation temperature of 15 K. Note that the results do not change within the error limits, even if we adopt the excitation temperature of 10 K.
The D/H ratio (i.e., what fraction of the hydrogen atom is replaced by the deuterium atom in the species of interest) is not the same as the abundance ratio, when the molecular species contains multiple equivalent H or D atoms. The D/H ratios translated from the derived abundance ratios are also shown in Table \ref{Dratio}.

The abundance ratios derived in this study are consistent with the previous reports, as shown in Table \ref{Dratio}.
Carbon-chain molecules (CCH and HC$_3$N) and standard species (HCN, HCO$^+$, N$_2$H$^+$, and HNC) show moderate D/H ratios of $\sim0.04$--$0.05$.
Note that the D/H ratios of HCO$^+$ and HNC can be regarded as an upper limit and lower limit, respectively. The H\ss{13}CO$^+$ and DNC lines could be optically thick, if we adopt the excitation temperature lower than 10 K. For overcoming this situation, using the double isotope species such as DN\ss{13}C to evaluate the D/H ratio would be preferable.
CH$_2$DCCH is statistically favored by a factor of 3, because there are three equivalent hydrogen atoms in the methyl group to be deuterated. Thus, the  [CH$_2$DCCH]/[CH$_3$CCH] ratio of 0.141 means a similar level of deuterium fractionation to the other carbon-chain species, when the statistical weight is considered.

The moderate D/H ratio observed in this study is consistent with the interpretation reported by \citet{s09b} and \citet{sy13}. In the WCCC sources, the starless-core phase would be shorter than the chemical timescale of the C-to-CO conversion ($\sim10^{6}$ yr). The starless-core phase is not long enough for deuterium transfer reactions and CO depletion onto dust grains to occur.  Thus, the D/H ratios should be relatively low in the WCCC sources \citep{s09b}. This is in contrast to the hot corino sources, which is thought to be a consequence of a longer cold starless-core phase with high degree of CO depletion.  In fact, the [CCD]/[CCH] ratio is reported to be 0.18 in the hot core source IRAS 16293-2422 \citep{vandishoeck95}, which is much higher than that in L1527 (0.037).

On the other hand, the D/H ratios of c-C$_3$D$_2$, D$_2$CO, and HDCS are found to be high in comparison with those of other species. This trend is seen in other low-mass star-forming regions, as described below. The [HDCS]/[H$_2$CS] ratio is 0.29, which is much higher than the abundance ratios of the other species. The D/H ratio is still as high as 0.14, in which the statistical weight of 2 due to the two equivalent hydrogen atoms is taken into account. The high [HDCS]/[H$_2$CS] ratio (0.29) is previously reported in the evolved starless core, Barnard 1, which is higher than the D/H ratios of the other molecules observed in this source \citep{marcelino05}. \citet{marcelino05} reproduced the high abundance ratio of HDCS in their steady-state gas-phase chemical model.

The derived [D$_2$CO]/[H$_2$CO] ratio of 0.014 is also high, because the D/H ratio ($\sqrt{\mathrm{[D_2CO]/[H_2CO]}}$) is as high as $0.12$. The high [D$_2$CO]/[H$_2$CO] ratio is often revealed in many low-mass star-forming regions including L1527 (e.g., \cite{rm07}; \cite{parise06}). \citet{parise06} reported the [D$_2$CO]/[H$_2$CO] ratio in L1527 (0.44), while \citet{rm07} reported the [D$_2$CO]/[H$_2$CO] ratio of 0.016. Our result is close to the result by \citet{rm07}. Several mechanisms both in the gas-phase and the solid-phase are proposed to account for the high D$_2$CO abundance. \citet{taquet12} modeled the high D/H ratio, considering the abstraction and substitution of H by D atom on grain surface (e.g., $\rm{H_2CO}+\rm{D}\to\rm{HDCO}+\rm{H}$; $\rm{HDCO}+\rm{D}\to\rm{D_2CO}+\rm{H}$).
\citet{roueff07} suggested that the D/H ratio of formaldehyde is related to the deuterium fractionation through CH$_3^+$ (i.e., $\mathrm{CH_3^+}+\mathrm{HD}\to\mathrm{CH_2D^+}+\mathrm{H_2}+390\ \mathrm{K}$) in their steady-state gas-phase model, because CH$_3$ produced from CH$_3^+$ reacts with O atoms to form formaldehyde.

The derived [c-C$_3$D$_2$]/[c-C$_3$HD] ratio of 0.11 in L1527 is higher than the [c-C$_3$HD]/[c-C$_3$H$_2$] ratio (0.044).
The [c-C$_3$HD]/[c-C$_3$H$_2$] ratio of 0.044 is even lower than those of other molecules, when the equivalent two H atoms in c-C$_3$H$_2$ are considered.
Since the abundance ratio of a doubly deuterated species, [XD$_2$]/[XHD], should be statistically 4 times lower than that of the singly deuterated one, [XHD]/[XH$_2$], this result means that c-C$_3$HD is more readily deuterated than c-C$_3$H$_2$.  On the other hand, \citet{spezzano13} reported that the [c-C$_3$D$_2$]/[c-C$_3$HD] ratio is comparable to the [c-C$_3$HD]/[c-C$_3$H$_2$] ratio in the starless cores L1544 and TMC-1C, although the errors of the ratios are large. They concluded that the observed D/H ratios can be explained by simple gas-phase reactions (e.g.,  the successive deuteration of c-C$_3$H$_2$ by the reaction with H$_2$D$^+$, D$_2$H$^+$, and D$_3^+$). However, the situation can be different in L1527, because the overabundance of c-C$_3$D$_2$ relative to c-C$_3$HD cannot be explained by considering only a single formation pathway. Hence, the result is puzzling, and chemical models may need to be revisited.

Note that the anomalous D/H ratios may also be related to the large beam size of the single-dish telescope.
Recently, Yoshida et al. (in preparation) have conducted high angular resolution observations of D$_2$CO and H$_2$CO in L1527 with ALMA, revealing their different distributions: the normal species mainly resides in the protostar position, whereas the deuterated species mainly resides in the outer envelope. The D/H ratios derived from the single-dish observations may therefore suffer from systematic errors caused by the simplified assumption of the same emitting region and the same excitation temperature. Thus, observations with high spatial resolution are needed to solve the origin of the high D/H ratios of some molecules in this source.

\subsubsection{\ss{13}C-substituted species}\label{sect13C}
In this survey, the \ss{13}C species of CO, C\ss{18}O, CN, HCO$^+$, DCO$^+$, HNC, HCN, CCH, c-C$_3$H$_2$, HC$_3$N, and H$_2$CO are detected. Among them, we derive the \ss{12}C/\ss{13}C ratios for the cases that the column densities of the corresponding normal species are reasonably determined without suffering from the high optical depth problem. The ratios for C\ss{18}O, CN, CCH, HC$_3$N, and c-C$_3$H$_2$ are summarized in Table \ref{13Cratio}. For molecules for which the rotation temperatures are not derived by the fit, the ratios are derived by assuming the excitation temperature of 15 K. Note that the results do not change within the error limits, even if the excitation temperature of 10 K is adopted.

Recent studies revealed that the \ss{12}C/\ss{13}C ratios of molecules produced from C$^+$ in the starless cores (TMC-1, L1521E, L1521B, and L134N) and L1527 are significantly higher than the elemental \ss{12}C/\ss{13}C ratio in the local interstellar medium (60--70; \cite{LL98}; \cite{milam05}). In L1527, CCH, HC$_3$N and c-C$_3$H$_2$ show dilution of \ss{13}C species except for HCC\ss{13}CN and c-CC\ss{13}CH$_2$ (\cite{s10b}; \cite{araki16}; \cite{taniguchi16b}; \cite{y15}). For CCH, the dilution is confirmed in this survey, where the column density of the normal species is derived from the optically-thin hyperfine components ($J=3/2$--$1/2$, $F=1$--$1$; $\tau=0.14$, and $J=3/2$--$1/2$, $F=1$--$0$; $\tau=0.13$). On the other hand, the derived \ss{12}C/\ss{13}C ratios of HC$_3$N are much less accurate than the reported values, because the sensitivity of this survey is not as good as the previous work \citep{araki16}. The derived \ss{12}C/\ss{13}C ratios of c-C$_3$H$_2$ are 1.5 times lower than those reported by \citet{y15}. This discrepancy would originate from the limited numbers of detected lines. \citet{y15} used 41 lines including optically thin lines in 1.3--3 mm bands to derive the column density of the normal species, while only 7 lines are detected in this survey. Moreover, some of them are relatively optically thick. The optically-thick lines suppress the column density of the normal species, which would make the \ss{12}C/\ss{13}C ratio low.

The dilution of \ss{13}C species was theoretically predicted by \citet{langer84}. In molecular clouds, the main reservoir of \ss{13}C is \ss{13}CO, and \ss{13}C$^+$ is produced by the reaction with He$^+$. However, \ss{13}C$^+$ can quickly go back to \ss{13}CO through the isotope exchange reaction,
\begin{eqnarray}
\mathrm{{}^{13}C^+ + CO\to C^+ + {}^{13}CO + 35\ K.}
\end{eqnarray}
Because this reaction is exothermic \citep{watson78}, the backward reaction is less likely to occur at low temperature conditions. This leads to the dilution of \ss{13}C$^+$, resulting in high \ss{12}C/\ss{13}C ratios of molecules produced from C$^+$. On the other hand, molecules produced directly from CO are expected to have the \ss{12}C/\ss{13}C ratio of 60--70, because \ss{13}CO is the main reservoir of \ss{13}C.
This phenomenon is already confirmed in the cold starless core TMC-1: various carbon-chain molecules show the dilution of \ss{13}C (Sakai et al.\ \yearcite{s07,s10b,s13}), while CH$_3$OH, which is formed through successive hydrogenation of CO on dust grains \citep{wk02}, shows the normal \ss{12}C/\ss{13}C ratio \citep{soma15}. Since the derived [C\ss{18}O]/[\ss{13}C\ss{18}O] ratio in L1527 is consistent with the elemental ratio within the error,
our study further confirms the dilution of \ss{13}C species of carbon-chain molecules in the protostellar core, L1527.

The above situation is different from a high [\ss{12}CO]/[\ss{13}CO] ratio caused by the isotope selective photodissociation (e.g., \cite{visser09}). This mechanism enhances the \ss{13}C$^+$ abundance, and subsequently decreases the \ss{12}C/\ss{13}C ratio in various molecules formed from \ss{13}C$^+$ (enrichment of the \ss{13}C species). This is opposite to our observational trend. Thus, this mechanism seems unimportant in this source. It is also suggested that a high [\ss{12}CO]/[\ss{13}CO] ratio in the gas phase is caused by the difference of the binding energy between \ss{13}CO and \ss{12}CO (\cite{smith15}; \cite{jorgensen18}). This can cause the lower abundance of \ss{13}C$^+$ relative to \ss{12}C$^+$, resulting in lower abundances of the \ss{13}C species of various molecules. However, the dilution of the \ss{13}C species are widely found even in cold clouds, and this mechanism seems less important than the mechanism mentioned in the above paragraph.

The derived [CN]/[\ss{13}CN] ratio $(61\pm17)$ is also comparable to the elemental ratio.
In diffuse clouds, dilution of \ss{13}C$^+$ is not significant due to high temperature. If CN produced in this earlier phase of molecular cloud evolution still remains, the \ss{12}C/\ss{13}C ratio of CN is close to $60$--$70$.
The derived \ss{12}C/\ss{13}C ratio of CN is consistent with the possible formation pathway of HC$_3$N: $\rm{C_2H_2}+\rm{CN}\to\rm{HC_3N}+\rm{H}$ (\cite{FO96}; Woon \& Herbst \yearcite{WH96}; \yearcite{WH97}; \cite{takano98}).  In the formation reaction of HC$_3$N, the C-N bond in CN is considered to be preserved.  Thus, the [HC$_3$N]/[HCC\ss{13}CN] ratio would be similar to the [CN]/[\ss{13}CN] ratio of 61. This expectation is now verified in L1527 by the [HC$_3$N]/[HCC\ss{13}CN] ratio of $49\pm15$ (this study) and $64.2\pm1.1$ \citep{araki16}.

\subsubsection{\ss{15}N-substituted species}
The \ss{15}N species of CN, HCN, and HNC are detected in this survey. The \ss{14}N/\ss{15}N ratios of the three species are consistent with one another within the error, as shown in Table \ref{15Nratio}. Sources where the \ss{14}N/\ss{15}N ratios of CN, HCN, and HNC are derived are limited to Barnard 1 and the protocluster OMC-2 FIR4. In both sources, the \ss{14}N/\ss{15}N ratios of the above three species are comparable to one another \citep{daniel13, kahane18}, and the ratios are similar to those derived in L1527 (Table \ref{15Nratio}). This result is consistent with the astrochemical models, which predict that the molecules bearing the nitrile functional group (-CN) have a common fractionation process (e.g., \cite{RC08}). On the other hand,  the \ss{14}N/\ss{15}N ratio of CN is reported to be significantly higher than that of HCN in the starless core L1544 (Hily-Blant et al. \yearcite{hily13a, hily13b}). This report contradicts with our result in L1527. Further observations are thus needed to solve this discrepancy.

\citet{araki16} reported that the \ss{14}N/\ss{15}N ratio of HC$_3$N is $338\pm12$ in L1527. If HC$_3$N is produced from the reaction mentioned above ($\rm{C_2H_2}+\rm{CN}\to\rm{HC_3N}+\rm{H}$), the \ss{14}N/\ss{15}N ratios of HC$_3$N and CN should be similar. On the other hand, the \ss{14}N/\ss{15}N ratio of CN ($230\pm80$) is derived to be lower than that of HC$_3$N. However, the error of this study is large, and is a target for future observations.

\subsection{Comparison with other sources}
\subsubsection{TMC-1}
In Figure \ref{N_tmc1}, the column densities of molecules derived in this study in L1527 are compared with those reported for TMC-1, which is the representative carbon-chain rich starless core. Basically, the column densities are well correlated between the two sources. We note the following trends.
\begin{enumerate}\raggedright
\item The column density ratios ($N_\mathrm{TMC-1}/N_\mathrm{L1527}$) of the nitrogen-bearing species (e.g., HC$_3$N and HC$_5$N) and the sulfur-bearing species (e.g., CCS and H$_2$CS) are higher than those of the hydrocarbons.
\item For C$_\mathrm{n}$H, longer chain molecules have relatively low $N_\mathrm{TMC-1}/N_\mathrm{L1527}$ ratios. For instance, the column densities of CCH and c/l-C$_3$H are higher than those in TMC-1, while the column densities of C$_4$H and C$_5$H are lower. This trend is more significant in longer chains as reported by \citet{s08b}. Note that \citet{araki17} reported that C$_7$H and C$_6$H$_2$ are the exception: relative abundances of C$_7$H and C$_6$H$_2$ in L1527 compared to TMC-1 are higher than those of C$_6$H and C$_4$H$_2$, respectively.
\end{enumerate}
The characteristics of WCCC, which are originally suggested by \citet{s08b}, are further confirmed.
In WCCC, carbon-chain molecules are efficiently regenerated from CH$_4$, which is sublimated from grain mantle after the onset of star formation.  Longer chains tend to be deficient in comparison with TMC-1, because they are produced more slowly than shorter chains. The formation timescale of N-bearing chains can also be long due to slow neutral-neutral reactions, suppressing the abundances of HC$_\mathrm{n}$N and other N-bearing species in WCCC sources relative to the carbon-chain molecules. S-bearing species can remain deficient, if sulfur is still depleted on dust grains.

As for COMs, the column density ratios relative to CH$_3$OH in L1527 and TMC-1 are compared in Table \ref{COMs}. The column densities in TMC-1 are taken from \citet{soma18}. The column density ratios in L1527 are almost comparable to those in TMC-1, suggesting that, unlike the hot corino case, the formation of COMs is inefficient. This is consistent with the WCCC mechanism (e.g., \cite{sy13}).

\subsubsection{IRAS 16293-2422}
Figure \ref{N_16293} shows the correlation plot of the column densities derived in L1527 and those in IRAS 16293-2422. This source is known to be a binary source, and the chemistry of each component has been studied by interferometoric observations (e.g., \cite{jorgensen16}). However, we here discuss the beam averaged column densities at the protostellar core scale (a few 1000 au) for a fair comparison with our line survey result for L1527 at a similar size scale. The beam-averaged column densities (beam size$\sim11''$--$28''$) of CH$_3$CHO, HCOOCH$_3$, and (CH$_3$)$_2$O in IRAS 16293-2422 are derived from the data reported by \citet{cazaux03} by using a least-squares method, under the assumption of the LTE condition. For other species, the beam-averaged column densities are derived from the spectral line survey observation in 0.9--3 mm bands (TIMASSS; \cite{caux11}) with the same method as for L1527 described in Section 3. We note the following trends.
\begin{enumerate}\raggedright
\item The column density ratios ($N_\mathrm{I16293}/N_\mathrm{L1527}$) of the carbon-chain related molecules, CCH, C$_4$H, HC$_3$N, and c-C$_3$H$_2$, are around 0.1--1.
\item On the other hand, the column density ratios of the CH$_3$OH, H$_2$CO, and CH$_3$CHO are higher than $\sim10$. Complex organic molecules such as HCOOCH$_3$ and (CH$_3$)$_2$O, which are characteristic species to hot corino sources, are not detected in L1527. The upper limits of HCOOCH$_3$ and (CH$_3$)$_2$O in L1527 suggest that these species are significantly less abundant than in IRAS 16293-2422 by two orders of magnitude or more.
Table \ref{COMs} shows the column density ratios of COMs relative to CH$_3$OH in L1527 and IRAS 16293-2422. The ratios of HCOOCH$_3$ and (CH$_3$)$_2$O relative to methanol in IRAS 16293-2422 are much higher than those derived toward L1527, indicating the efficient production of COMs in the hot corino. The ratio of the column density of H$_2$CO in L1527 relative to that in IRAS 16293-2422 is higher than the corresponding ratios of COMs (CH$_3$CHO, HCOOCH$_3$, and (CH$_3$)$_2$O) relative to CH$_3$OH. The high abundance of H$_2$CO relative to CH$_3$OH in L1527 is consistent with the fast contraction scenario in WCCC, because the timescale for the formation of CH$_3$OH on dust grains is longer than that of H$_2$CO \citep{taquet12}.  On the other hand, CH$_3$CHO shows the similar ratio among the 3 sources shown in Table \ref{COMs}. This result implies that CH$_3$CHO may not be enhanced by the hot corino chemistry.
\item The column density ratios of the S-bearing species are around 10--100 except for CCS. CCS is efficiently produced during chemically young cold-core phase \citep{suzuki92}, and is not enhanced by the activity of the central protostar \citep{hirota10}. The faster contraction timescale in L1527 may result in the higher abundance of 'remnant' CCS in comparisom with other S-bearing species.
\end{enumerate}
As shown in Figure \ref{N_16293}, the $N_\mathrm{I16293}/N_\mathrm{L1527}$ ratios range from $10^{2}$--$10^{3}$ (COMs) to 0.1 (carbon-chain molecules).
Chemical difference with such a high dynamic range of the column density ratio ($\gtsim3$ orders of magnitude) is noteworthy. For instance, \citet{watanabe12} conducted a spectral line survey toward the Class 0--I protostar R$\;$CrA$\;$IRS7B, revealing a higher abundance of CCH and lower abundances of CH$_3$OH and SO$_2$ in comparison with IRAS 16293-2422. The dynamic range of the column density ratios is 2--3 orders of magnitude, which is lower than the case between L1527 and IRAS 16293-2422. \citet{watanabe12} concluded that R$\;$CrA$\;$IRS7B is a source with a mixture of hot corino chemistry and WCCC or a source under a strong influence of the external UV radiation. The higher dynamic range of the column density ratios shown in Figure \ref{N_16293} further confirms that the hot corinos such as IRAS 16293-2422 and the WCCC sources such as L1527 are the two distinct cases in chemical composition.

The full chemical composition of sources can be characterized without any preconception only by an unbiased spectral line survey with high sensitivity that can detect not only the major species but also the minor species such as the COMs and carbon-chain molecules. Since unbiased spectral line survey observations toward low-mass star-forming regions have been carried out only toward a few representative sources such as IRAS 16293-2422, R CrA IRS7B, and L1527, those toward other low-mass star-forming regions are of particular importance to understand the origin of the chemical diversity.
\bigskip
\begin{ack}
The authors thank Cecilia Ceccarelli and Bertrand Lefloch for valuable discussions, particularly for communication of the ASAI data prior to publication, and Sheng-Yuan Liu for valuable comments. The authors are grateful to the Nobeyama Radio Observatory
(NRO) staff for great support in the observations with the 45m telescope.
The Nobeyama Radio Observatory is a branch of the National Astronomical Observatory of Japan, National Institutes of Natural Sciences.
K. Y. acknowledges the RIKEN Junior Research Associate Program and the JSPS fellowship.
This study is supported by Grant-in-Aids from the Ministry of Education, Culture, Sports, Science, and Technology of Japan (Nos. 21740132, 23740142, 25400223, 16H03964, and 25108005).
N. S. and S. Y. acknowledge financial support by JSPS and MAEE under the Japan-France integrated action program.
\end{ack}
\appendix
\section*{Observed frequencies of the $N=1$--$0$ transition of CCD}
Some of the hyperfine components of the CCD $N=1$--$0$ transition are observed with slightly different frequencies from those reported in CDMS, as can be seen in Figure \ref{spect_4mm} (e.g., the top middle panel). The observed frequencies are summarized in Table \ref{tab_CCD}. The rest frequencies reported in CDMS may be inaccurate, because they are calculated based on the laboratory measurements of the higher $N$ transitions \citep{Bogey85,Vrtilek85}.
The accurate rest frequencies need to be determined in the future laboratory experiment.

\clearpage
\begin{table}
\tbl{Molecules detected in this study}{
\begin{tabular}{l}
\hline
\multicolumn{1}{c}{Carbon-chain molecules and the isomers} \\
\hline
CCH, \ss{13}CCH, C\ss{13}CH, CCD, C$_2$O$^*$, CCS, \\
C$_3$O, C$_3$N, l-C$_3$H, l-C$_3$D$^*$, c-C$_3$H, c-C$_3$D,\\
c-C$_3$H$_2$, c-\ss{13}CCCH$_2$, c-CC\ss{13}CH$_2$, c-C$_3$HD, c-C$_3$D$_2$, \\
l-C$_3$H$_2$, C$_3$S$^*$, CH$_3$CCH, CH$_2$DCCH, CH$_3$CCD$^*$, \\
HCCNC, HC$_3$N, DC$_3$N, H\ss{13}CCCN, HC\ss{13}CCN, HCC\ss{13}CN,\\
C$_4$H, C$_4$D$^*$, CC\ss{13}CCH$^*$, C$_4$H$_2$, C$_5$H, HC$_5$N, C$_6$H$^*$\\
\hline
\multicolumn{1}{c}{Complex organic molecules} \\
\hline
CH$_3$OH, CH$_2$DOH$^*$, CH$_3$CHO, c-H$_2$C$_3$O, HCCCHO, C$_2$H$_3$CN$^*$\\
\hline
\multicolumn{1}{c}{Other species} \\
\hline
CO, \ss{13}CO, C\ss{18}O, C\ss{17}O, \ss{13}C\ss{18}O, CS, C\ss{34}S, C\ss{33}S, \\
CN,  \ss{13}CN, C\ss{15}N, SO, HNC, HN\ss{13}C, H\ss{15}NC, DNC,\\
HCN, H\ss{13}CN, HC\ss{15}N, DCN, N$_2$H$^+$, N$_2$D$^+$, \\
HCO$^{+}$, H\ss{13}CO$^{+}$, HC\ss{18}O$^{+}$, DCO$^{+}$, D\ss{13}CO$^{+}$ \\
HCS$^+$, HCO, HNCO, NH$_2$D, H$_2$\ss{13}CO, D$_2$CO, \\
H$_2$CS, HDCS, CH$_2$CN, CH$_2$CO, CHDCO, HCO$_2^+$$^*$, HCNO$^*$\\
\hline
\end{tabular}}\label{tab_detected}%
\begin{tabnote}
\footnotemark[*] Tentative detection\\
\end{tabnote}
\end{table}

\begin{longtable}{cp{1.6cm}p{3.5cm}p{1cm}p{1.5cm}p{1.5cm}p{1.5cm}p{1.8cm}p{.8cm}}
\caption{Line parameters observed in the 3 mm band\footnotemark[$*$]. }\label{line_parameter_3mm}
\hline 
Frequency & \centering{Molecule} & \centering{Transition} & \centering{$E_\mathrm{u}$} &
$\Delta V_\mathrm{LSR}$\footnotemark[$*$] & $T_\mathrm{mb}$\footnotemark[$*$] &
$\Delta v$\footnotemark[$*$] & $\int T_\mathrm{mb}dv$ & rms \\
(GHz) & & & \centering{(cm$^{-1}$)} & (km s$^{-1}$) & (K) & (km s$^{-1}$) & (K km s$^{-1}$) & (mK) \\
\hline
\endhead
\hline
\endfoot
\hline
\multicolumn{9}{@{}l@{}}{\hbox to 0pt{\parbox{180mm}{\footnotesize
\par\noindent
\footnotemark[$*$] Obtained from the Gaussian fit. $\Delta V_\mathrm{LSR}$ represents the velocity shift from $V_\mathrm{LSR}$ of 5.85 km s$^{-1}$.  $\Delta v$ represents the FWHM line width. ``$\ldots$'' shows that the fit is not successful due to the weak intensity or the insufficient resolution.\par\noindent
\footnotemark[$\dagger$] Less than $3\sigma$ detection.\par\noindent
\footnotemark[$\ddagger$] Less than $4\sigma$ detection.\par\noindent
\footnotemark[$\S$] Blended.\par\noindent
\footnotemark[$\|$] The line is not detected. Three times the rms noise is used to derive the upper limit of the column density by assuming the line width of 1 km s$^{-1}$.}\hss}}
\endlastfoot79.812333&$\mathrm{c\mathchar`-C_3HD}$&$2_{1,2}$--$1_{0,1}$&4.1&$-0.059(7)$&$0.477(6)$&$1.164(18)$&$0.600(9)$&6.7\\
79.876710&$\mathrm{HC_5N}$&$30$--$29$&41.3&$-0.04(2)$&$0.118(5)$&$1.25(6)$&$0.170(5)$&4.8\\
80.046678&$\mathrm{C_4H_2}$&$9_{1,9}$--$8_{1,8}$&22.7&$-0.10(3)$&$0.145(6)$&$1.40(7)$&$0.239(9)$&6.4\\
80.047537&$\mathrm{c\mathchar`-CC{}^{13}CH_2}$&$2_{0,2}$--$1_{1,1}$&4.4&$-0.07(2)$&$0.164(6)$&$1.26(5)$&$0.229(8)$&7.4\\
80.076652&$\mathrm{CH_2CO}$&$4_{1,4}$--$3_{1,3}$&15.7&$0.02(5)$&$0.088(7)$&$1.36(11)$&$0.140(11)$&8.2\\
80.383887&$\mathrm{C_4H_2}$&$9_{0,9}$--$8_{0,8}$&13.4&$-0.22(2)$&$0.101(4)$&$1.19(5)$&$0.135(5)$&4.2\\
80.412848&$\mathrm{l\mathchar`-C_3H}$&${}^2\Pi_{1/2}$, $J=$7/2--5/2, $F=$3--2, $f$&19.4&$0.0(2)$&$0.015(4)$\footnotemark[$\dagger$]&$1.6(5)$&$0.029(9)$&6.1\\
80.420646&$\mathrm{l\mathchar`-C_3H}$&${}^2\Pi_{1/2}$, $J=$7/2--5/2, $F=$4--3, $f$&19.4&$-0.1(2)$&$0.014(4)$\footnotemark[$\dagger$]&$1.6(5)$&$0.030(6)$&5.0\\
\end{longtable}

\begin{longtable}{cp{1.6cm}p{3.5cm}p{1cm}p{1.5cm}p{1.5cm}p{1.5cm}p{1.8cm}p{.8cm}}
\caption{Line parameters of the supplementary observation in the 4 mm band.}\label{line_parameter_4mm}
\hline 
Frequency & \centering{Molecule} & \centering{Transition} & \centering{$E_\mathrm{u}$} &
$\Delta V_\mathrm{LSR}$\footnotemark[$*$] & $T_\mathrm{mb}$\footnotemark[$*$] &
$\Delta v$\footnotemark[$*$] & $\int T_\mathrm{mb}dv$ & rms \\
(GHz) & & & \centering{(cm$^{-1}$)} & (km s$^{-1}$) & (K) & (km s$^{-1}$) & (K km s$^{-1}$) & (mK) \\
\hline
\endhead
\hline
\endfoot
\hline
\multicolumn{9}{@{}l@{}}{\hbox to 0pt{\parbox{180mm}{\footnotesize
\par\noindent
\footnotemark[$*$] Obtained from the Gaussian fit. $\Delta V_\mathrm{LSR}$ represents the velocity shift from $V_\mathrm{LSR}$ of 5.85 km s$^{-1}$. $\Delta v$ represents the FWHM line width.\par\noindent
\footnotemark[$\dagger$] Less than $3\sigma$ detection.\par\noindent
\footnotemark[$\ddagger$] Less than $4\sigma$ detection.}\hss}}
\endlastfoot
70.733206&$\mathrm{D{}^{13}CO+}$&$1$--$0$&2.4&$0.12(3)$&$0.28(3)$&$0.53(7)$&$0.153(17)$&32.4\\
71.024788&$\mathrm{H_2{}^{13}CO}$&$1_{0,1}$--$0_{0,0}$&2.4&$0.13(5)$&$0.18(3)$&$0.63(11)$&$0.124(17)$&30.2\\
72.039312&$\mathrm{DCO^+}$&$1$--$0$&2.4&$0.061(3)$&$3.31(3)$&$0.654(6)$&$2.308(17)$&30.3\\
72.101715&$\mathrm{CCD}$&$N=$1--0, $J=$3/2--1/2, $F=$3/2--3/2&2.4&$-0.27(4)$&$0.20(3)$&$0.58(9)$&$0.147(16)$&30.0\\
72.107700&$\mathrm{CCD}$&$N=$1--0, $J=$3/2--1/2, $F=$5/2--3/2&2.4&$-0.035(13)$&$0.61(3)$&$0.60(3)$&$0.395(16)$&29.2\\
72.109114&$\mathrm{CCD}$&$N=$1--0, $J=$3/2--1/2, $F=$1/2--1/2&2.4&$0.40(3)$&$0.24(3)$&$0.43(6)$&$0.117(14)$&29.8\\
72.112399&$\mathrm{CCD}$&$N=$1--0, $J=$3/2--1/2, $F=$3/2--1/2&2.4&$0.51(2)$&$0.33(3)$&$0.48(6)$&$0.144(15)$&30.2\\
72.187704&$\mathrm{CCD}$&$N=$1--0, $J=$1/2--1/2, $F=$3/2--3/2&2.4&$-0.03(2)$&$0.36(3)$&$0.50(6)$&$0.165(20)$&39.8\\
\end{longtable}
\clearpage
\begin{table}
\tbl{Derived column densities, rotation temperatures and optical depths of the lines used for the evaluation\footnotemark[$*$]}{
\begin{tabular}{lcccrrlccc}
\hline
\multicolumn{1}{l}{Molecule} & $N$ (cm$^{-2}$) & $T_\mathrm{rot}$ (K) & $\tau$ range\footnotemark[$\dagger$]
&&& \multicolumn{1}{l}{Molecule} & $N$ (cm$^{-2}$) & $T_\mathrm{rot}$ (K) & $\tau$ range\footnotemark[$\dagger$]\\
\hline
SO&$1.1(9)\times10^{13}$&6(2)&$<0.21$&&&C$_5$H&$5.3(11)\times10^{11}$&33(26)&$<0.002$\\
CCS&$6(2)\times10^{12}$&6.1(8)&$<0.066$&&&HC$_3$N&$1.22(6)\times10^{13}$&25(4)&$<0.16$\\
C$_3$O&$1.8(11)\times10^{11}$&11(5)&$<0.034$&&&H\ss{13}CCCN&$1.4(4)\times10^{11}$&25 (fix)\footnotemark[$\ddagger$]&$<0.002$\\
C$_3$N&$6(7)\times10^{11}$&11(9)&$<0.009$&&&HC\ss{13}CCN&$2.4(3)\times10^{11}$&25 (fix)\footnotemark[$\ddagger$]&$<0.0028$\\
l-C$_3$H&$3.5(2)\times10^{12}$&11.5(12)&$<0.067$&&&HCC\ss{13}CN&$2.5(8)\times10^{11}$&25 (fix)\footnotemark[$\ddagger$]&$<0.003$\\
c-C$_3$H&$1.4(5)\times10^{13}$&15(6)&$<0.034$&&&DC$_3$N&$4.7(12)\times10^{11}$&25 (fix)\footnotemark[$\ddagger$]&$<0.015$\\
c-C$_3$H$_2$&$4.4(5)\times10^{13}$&8.9(6)&$<0.83$&&&HC$_5$N&$2.4(18)\times10^{12}$&22(7)&$<0.005$\\
c-\ss{13}CCCH$_2$&$2.18(16)\times10^{11}$&8.9 (fix)\footnotemark[$\ddagger$]&$<0.013$&&&CH$_3$CCH&$7.0(6)\times10^{13}$&14.2(15)&$<0.070$\\
c-CC\ss{13}CH$_2$&$1.07(16)\times10^{12}$&8.9 (fix)\footnotemark[$\ddagger$]&$<0.058$&&&CH$_2$DCCH&$9.83(13)\times10^{12}$&14.2 (fix)\footnotemark[$\ddagger$]&$<0.008$\\
c-C$_3$HD&$1.9(2)\times10^{12}$&8.9 (fix)\footnotemark[$\ddagger$]&$<0.11$&&&CH$_3$OH&$2.3(10)\times10^{13}$&13(8)&$<0.030$\\
c-C$_3$D$_2$&$2.2(8)\times10^{11}$&8.9 (fix)\footnotemark[$\ddagger$]&0.006&&&CH$_3$CN&$2.54(15)\times10^{11}$&21(2)&$<0.005$\\
l-C$_3$H$_2$&$1.23(7)\times10^{12}$&9.2(8)&$<0.049$&&&CH$_2$CN&$2.2(4)\times10^{12}$&8.6(19)&$<0.010$\\
C$_4$H&$2.32(12)\times10^{14}$&12.3(4)&$<0.26$&&&CH$_2$CO&$7(3)\times10^{12}$&27(23)&$<0.008$\\
C$_4$H$_2$&$1.9(2)\times10^{12}$&12.7(11)&$<0.017$&&&HCCCHO&$2.4(3)\times10^{12}$&16(13)&$<0.004$\\
\hline
\end{tabular}}\label{tab_NT}%
\begin{tabnote}
\footnotemark[$*$] The numbers in parentheses represent the errors in units of last significant digits.\\
\footnotemark[$\dagger$] The range of the optical depths of the lines used in the analysis.\\
\footnotemark[$\ddagger$] The rotation temperature is assumed to be the same as that derived for the normal species.\\
\end{tabnote}
\end{table}

\clearpage
\begin{table}
\tbl{Derived column densities and optical depths of the lines used for the evaluation\footnotemark[$*$]}{
\begin{tabular}{lccccrlcccc}
\hline
\multicolumn{1}{c}{Molecule} & \multicolumn{2}{c}{$T_\mathrm{rot}=10$ K} & \multicolumn{2}{c}{$T_\mathrm{rot}=15$ K} &&
\multicolumn{1}{c}{Molecule} & \multicolumn{2}{c}{$T_\mathrm{rot}=10$ K} & \multicolumn{2}{c}{$T_\mathrm{rot}=15$ K} \\
\cline{2-3}\cline{4-5}\cline{8-9}\cline{10-11}
& $N$ (cm$^{-2}$) & $\tau$ range & $N$ (cm$^{-2}$) & $\tau$ range &&
& $N$ (cm$^{-2}$) & $\tau$ range & $N$ (cm$^{-2}$) & $\tau$ range \\
\hline
CO&($4.4(16)\times10^{18}$)\footnotemark[$\dagger$]&--&($4.0(10)\times10^{18}$)\footnotemark[$\dagger$]&--&&DCN&$2.6(5)\times10^{12}$&$<0.11$&$3.1(6)\times10^{12}$&$<0.061$\\
\ss{13}CO&($5.1(13)\times10^{16}$)\footnotemark[$\ddagger$]&--&($5.9(15)\times10^{16}$)\footnotemark[$\ddagger$]&--&&N$_2$H$^+$&$5.8(3)\times10^{12}$&$<0.20$&$6.6(13)\times10^{12}$&$<0.11$\\
C\ss{18}O&$8(3)\times10^{15}$&1.07&$7.1(18)\times10^{15}$&0.48&&N$_2$D$^+$&$3.0(6)\times10^{11}$&$<0.010$&$3.6(7)\times10^{11}$&$<0.006$\\
C\ss{17}O&$1.90(4)\times10^{15}$&$<0.16$&$2.1(4)\times10^{15}$&$<0.088$&&CCH\footnotemark[$\|$]&$1.4(3)\times10^{15}$&$<0.25$&$1.5(3)\times10^{15}$&$<0.14$\\
\ss{13}C\ss{18}O&$9(2)\times10^{13}$&0.011&$1.1(3)\times10^{14}$&0.007&&\ss{13}CCH&$6.3(10)\times10^{12}$&$<0.007$&$7.5(15)\times10^{12}$&$<0.004$\\
CS&($3.1(6)\times10^{13}$)\footnotemark[$\S$]&--&($3.4(7)\times10^{13}$)\footnotemark[$\S$]&--&&C\ss{13}CH&$9.4(19)\times10^{12}$&$<0.009$&$1.1(2)\times10^{13}$&$<0.005$\\
\hline
\end{tabular}}\label{tab_N_fix}%
\begin{tabnote}
\footnotemark[$*$] The numbers in parentheses represent the errors in units of last significant digits.\\
\footnotemark[$\dagger$] The spectral line of C\ss{18}O is used for evaluation of the column density of CO, where the \ss{16}O/\ss{18}O ratio is assumed to be 560 \citep{WR94}.\\
\footnotemark[$\ddagger$] The spectral line of \ss{13}C\ss{18}O is used for evaluation of the column density of C\ss{18}O, where the \ss{16}O/\ss{18}O ratio is assumed to be 560 \citep{WR94}.\\
\footnotemark[$\S$] The spectral lines of the \ss{34}S species are used for evaluation of the column density, where the \ss{32}S/\ss{34}S ratio is assumed to be 22 \citep{chin96}.\\
\footnotemark[$\|$] Two optically-thin hyperfine components are used for evaluation of the column density.\\
\footnotemark[$\#$] The spectral lines of the \ss{13}C species are used for evaluation of the column density, where the \ss{12}C/\ss{13}C ratio is assumed to be 60 \citep{LL98}.\\
\footnotemark[$**$] No lines are observed in this survey.\\
\end{tabnote}
\end{table}

\clearpage
\begin{table}
\tbl{Comparison of column densities derived by the non-LTE and LTE analysis \footnotemark[$*$]}{
\begin{tabular}{lcc}
\hline
\multicolumn{1}{c}{Molecule} & \multicolumn{2}{c}{$N$ (cm$^{-2}$)} \\
\cline{2-3}
&Non-LTE\footnotemark[$\dagger$]&LTE\footnotemark[$\ddagger$]\\
\hline
C\ss{18}O&(0.7--1.0)$\times10^{16}$&\cd{7.1}{18}{15}\\
CS&(1.4--2.0)$\times10^{13}$&\cd{3.4}{7}{13}\\
SO&(4.0--6.5)$\times10^{12}$&\cd{1.1}{9}{13}\\
HC\ss{18}O$^+$&(3.5--7.5)$\times10^{11}$&\cd{5.7}{11}{11}\\
HC\ss{15}N&(1.5--2.6)$\times10^{11}$&\cd{2.6}{5}{11}\\
H\ss{13}CN&(0.7--1.4)$\times10^{11}$&\cd{1.1}{2}{11}\\
CCH\footnotemark[$\|$]&(1.2--2.2)$\times10^{15}$&\cd{1.5}{3}{15}\\
CH$_3$OH\footnotemark[$\#$]&(2.3--3.4)$\times10^{13}$&\cd{2.3}{10}{13}\\
\hline
\end{tabular}}\label{tab_NLTE}%
\begin{tabnote}
\footnotemark[$*$] The H$_2$ density and the kinetic temperature are set to be $3\times10^5$--$3\times10^6\ \mathrm{cm^{-3}}$ and $10$--$30$ K, respectively.\\
\footnotemark[$\dagger$] Calculated with RADEX \citep{vdT07}. The collisional rate coefficients are taken from \citet{yang10} for C\ss{18}O, \citet{lique06} for CS and SO, \citet{flower99} for HC\ss{18}O$^+$, \citet{gt74} for HC\ss{15}N and H\ss{13}CN, \citet{spielfiedel12} for CCH, and \citet{rf10} for CH$_3$OH. \\
\footnotemark[$\ddagger$] The numbers in parentheses represent the errors in units of last significant digits. For molecules listed in Table \ref{tab_N_fix}, the column density derived with $T_\mathrm{rot}=15$ K is shown.\\
\footnotemark[$\|$] Two optically-thin hyperfine components are used.\\
\footnotemark[$\#$] The H$_2$ density is restricted to $3\times10^5$--$1\times10^6\ \mathrm{cm^{-3}}$ to reproduce the absorption feature of $3_{1,3}$--$4_{1,4},\;$A$^+$ line.\\
\end{tabnote}
\end{table}

\clearpage
\begin{table}
\tbl{Fractinal abundance relative to H$_2$, calculated from Tables \ref{tab_NT} and \ref{tab_N_fix}.\footnotemark[$*$]}{
\begin{tabular}{lccclccclc}
\hline
\multicolumn{1}{l}{Molecule} & $X$  &&
& \multicolumn{1}{l}{Molecule} & $X$ &&
& \multicolumn{1}{l}{Molecule} & $X$ \\
\hline
CO&($4.8(16)\times10^{-5}$)\footnotemark[$\dagger$]&&&N$_2$H$^+$&$8(2)\times10^{-11}$&&&HCC\ss{13}CN&$3.0(11)\times10^{-12}$\\
\ss{13}CO&($7(2)\times10^{-7}$)\footnotemark[$\ddagger$]&&&N$_2$D$^+$&$4.4(12)\times10^{-12}$&&&DC$_3$N&$5.7(18)\times10^{-12}$\\
C\ss{18}O&$9(3)\times10^{-8}$&&&CCH&$1.9(5)\times10^{-8}$&&&HC$_5$N&$3(2)\times10^{-11}$\\
\ss{13}C\ss{18}O&$1.3(4)\times10^{-9}$&&&\ss{13}CCH&$9(3)\times10^{-11}$&&&c-H$_2$C$_3$O&$5.5(17)\times10^{-12}$\\
CS&($4.2(12)\times10^{-10}$)\footnotemark[$\#$]&&&C\ss{13}CH&$1.4(4)\times10^{-10}$&&&H$_2$CO&($1.9(6)\times10^{-9}$)\footnotemark[$\#$,$**$]\\
C\ss{34}S&$1.9(5)\times10^{-11}$&&&CCD&$6.9(19)\times10^{-10}$&&&H$_2$\ss{13}CO&$3.1(10)\times10^{-11}$\\
C\ss{33}S&$2.8(10)\times10^{-12}$&&&HCS$^+$&$5.2(15)\times10^{-12}$&&&D$_2$CO&$2.7(11)\times10^{-11}$\\
CN&$3.2(9)\times10^{-9}$&&&HCO&$9(2)\times10^{-11}$&&&H$_2$CS&$3.5(9)\times10^{-11}$\\
\ss{13}CN&$5.2(14)\times10^{-11}$&&&CCS&$7(3)\times10^{-11}$&&&HDCS&$10(3)\times10^{-12}$\\
C\ss{15}N&$1.4(4)\times10^{-11}$&&&C$_3$O&$2.2(14)\times10^{-12}$&&&NH$_2$D&$1.8(17)\times10^{-11}$\\
SO&$1.3(11)\times10^{-10}$&&&l-C$_3$H&$4.3(9)\times10^{-11}$&&&HCCNC&$3.8(14)\times10^{-12}$\\
HCO$^+$&($3.8(11)\times10^{-9}$)\footnotemark[$\#$]&&&c-C$_3$H&$1.7(7)\times10^{-10}$&&&HNCO&$3.2(7)\times10^{-11}$\\
H\ss{13}CO$^+$&$6.3(19)\times10^{-11}$&&&c-C$_3$H$_2$&$5.3(12)\times10^{-10}$&&&CH$_3$CHO&$2.0(7)\times10^{-11}$\\
HC\ss{18}O$^+$&$6.9(19)\times10^{-12}$&&&c-\ss{13}CCCH$_2$&$2.6(6)\times10^{-12}$&&&CH$_3$CCH&$8.5(18)\times10^{-10}$\\
DCO$^+$&($2.2(7)\times10^{-10}$)\footnotemark[$\#$]&&&c-CC\ss{13}CH$_2$&$1.3(3)\times10^{-11}$&&&CH$_2$DCCH&$1.2(2)\times10^{-10}$\\
D\ss{13}CO$^+$&$3.6(11)\times10^{-12}$&&&c-C$_3$HD&$2.4(5)\times10^{-11}$&&&CH$_3$OH&$2.8(13)\times10^{-10}$\\
HNC&($1.4(4)\times10^{-9}$)\footnotemark[$\#$]&&&c-C$_3$D$_2$&$2.7(11)\times10^{-12}$&&&CH$_3$CN&$3.1(6)\times10^{-12}$\\
HN\ss{13}C&$2.3(7)\times10^{-11}$&&&l-C$_3$H$_2$&$1.5(3)\times10^{-11}$&&&CH$_2$CN&$2.6(7)\times10^{-11}$\\
H\ss{15}NC&$4.5(13)\times10^{-12}$&&&C$_4$H&$2.8(6)\times10^{-9}$&&&CH$_2$CO&$9(4)\times10^{-11}$\\
DNC&$6.3(19)\times10^{-11}$&&&C$_4$H$_2$&$2.4(5)\times10^{-11}$&&&HCCCHO&$3.0(7)\times10^{-11}$\\
HCN&($8(2)\times10^{-10}$)\footnotemark[$\#$]&&&C$_5$H&$6.4(18)\times10^{-12}$&&&OCS&$<2.3\times10^{-11}$\\
H\ss{13}CN&$1.3(4)\times10^{-11}$&&&HC$_3$N&$1.5(3)\times10^{-10}$&&&HCOOCH$_3$&$<1.2\times10^{-10}$\\
HC\ss{15}N&$3.1(9)\times10^{-12}$&&&H\ss{13}CCCN&$1.8(6)\times10^{-12}$&&&(CH$_3$)$_2$O&$<6.1\times10^{-11}$\\
DCN&$3.8(11)\times10^{-11}$&&&HC\ss{13}CCN&$2.9(7)\times10^{-12}$&&&&\\
\hline
\end{tabular}}\label{tab_X}%
\begin{tabnote}
\footnotemark[$*$] The numbers in parentheses represent the errors in units of last significant digits. The column density of $\mathrm{H_2}$ is evaluated from that of C${}^{17}$O to be $(8.2\pm1.6)\times10^{22}$ cm$^{-2}$. For molecules listed in Table \ref{tab_N_fix}, the column density derived with $T_\mathrm{rot}=15$ K is assumed.\\
\footnotemark[$\dagger$,$\ddagger$,$\|$,$\#$,$**$] See the footnotes in Table \ref{tab_N_fix}.
\end{tabnote}
\end{table}

\clearpage
\begin{table}
\tbl{D/H ratios\footnotemark[$*$]}{
\begin{tabular}{lccc}
\hline
\multicolumn{1}{c}{Ratios} & \multicolumn{2}{c}{This work} & Previous works\\
\cline{2-4}
& Abundance ratio & D/H ratio\footnotemark[$\dagger$] & Abundance ratio\\
\hline
D\ss{13}CO$^+$/H\ss{13}CO$^+$ & 0.057(19) && 0.048\footnotemark[$\ddagger$]\\
N$_2$D$^+$/N$_2$H$^+$ & 0.054(15) && 0.06(1)\footnotemark[$\S$]\\
DCN/HCN & 0.048(13)\footnotemark[$\|$] && 0.037(8)\footnotemark[$\|$,$\#$] \\
DNC/HNC & 0.045(14)\footnotemark[$\|$] && 0.046(4)\footnotemark[$\|$,$**$]\\
CCD/CCH & $0.037(10)$ && --\\
DC$_3$N/HC$_3$N & 0.039(10) && 0.0370(7)\footnotemark[$\dagger\dagger$] \\
c-C$_3$HD/c-C$_3$H$_2$ & 0.044(7) & 0.022(4) & 0.071(23)\footnotemark[$\ddagger\ddagger$]\\
c-C$_3$D$_2$/c-C$_3$H$_2$ & 0.0050(18) & 0.071(13) & --\\
c-C$_3$D$_2$/c-C$_3$HD & 0.11(4) & 0.23(8) & -- \\
CH$_2$DCCH/CH$_3$CCH & 0.141(12) & 0.047(4) & -- \\
HDCS/H$_2$CS & 0.29(10) & 0.14(5) & -- \\
D$_2$CO/H$_2$CO & 0.014(6)\footnotemark[$\|$] & 0.12(3) & 0.016(5)\footnotemark[$\S$,$\|$], $0.44^{+0.60}_{-0.29}$\footnotemark[$\S\S$]\\
\hline
\end{tabular}}\label{Dratio}%
\begin{tabnote}
\footnotemark[$*$] The numbers in parentheses represent the errors in units of the last significant digits.\\
\footnotemark[$\dagger$] The D/H ratio means the abundance ratio for the single deuteration, which is corrected for the statistical weight caused by molecular symmetry \citep{persson18}.\\
\footnotemark[$\ddagger$] The [DCO$^+$]/[HCO$^+$] ratio reported by \citet{jorgensen04}, where the column density of HCO$^+$ is derived from the H${}^{13}$CO$^+$ lines by assuming the \ss{12}C/\ss{13}C ratio of 60.\\
\footnotemark[$\S$] \citet{rm07}\\
\footnotemark[$\|$] The column density of the normal species is derived from the \ss{13}C species, where the \ss{12}C/\ss{13}C ratio is assumed to be 60.\\
\footnotemark[$\#$] \citet{roberts02}\\
\footnotemark[$**$] \citet{hirota01}\\
\footnotemark[$\dagger\dagger$] \citet{araki16}\\
\footnotemark[$\ddagger\ddagger$] \citet{s09b}\\
\footnotemark[$\S\S$] \citet{parise06}\\
\end{tabnote}
\end{table}

\clearpage
\begin{table}
\tbl{\ss{12}C/\ss{13}C ratios\footnotemark[$*$]}{
\begin{tabular}{lcc}
\hline
\multicolumn{1}{c}{Ratios} & This work & Previous works \\
\hline
[C\ss{18}O]/[\ss{13}C\ss{18}O] & 70(20) & -- \\
$[$CN]/[\ss{13}CN] & 61(17) & -- \\
$[$CCH]/[\ss{13}CCH] & 210(60) & $>135$\footnotemark[$\dagger$] \\
$[$CCH]/[C\ss{13}CH] & 140(40) & $>80$\footnotemark[$\dagger$] \\
$[$HC$_3$N]/[H\ss{13}CCCN] & 85(22) & 86.4(16)\footnotemark[$\ddagger$] \\
$[$HC$_3$N]/[HC\ss{13}CCN] & 51(7) & 85.4(17)\footnotemark[$\ddagger$] \\
$[$HC$_3$N]/[HCC\ss{13}CN] & 49(15) & 64.2(11)\footnotemark[$\ddagger$]\\
$[$c-C$_3$H$_2$]/[c-\ss{13}CCCH$_2$] & 200(30) & 310(80)\footnotemark[$\S$]\\ 
$[$c-C$_3$H$_2$]/[c-CC\ss{13}CH$_2$] & 41(8) & 61(11)\footnotemark[$\S$] \\
\hline
\end{tabular}}\label{13Cratio}%
\begin{tabnote} 
\footnotemark[$*$] The numbers in parentheses represent the errors in units of the last significant digits.\\
\footnotemark[$\dagger$] \citet{s10b}\\
\footnotemark[$\ddagger$] \citet{araki16}\\
\footnotemark[$\S$] \citet{y15}\\
\end{tabnote}
\end{table}
\clearpage
\begin{table}
\tbl{\ss{14}N/\ss{15}N ratios\footnotemark[$*$]}{
\begin{tabular}{lccc}
\hline
\multicolumn{1}{c}{Ratios} & L1527\footnotemark[$\dagger$] & B1\footnotemark[$\ddagger$] & OMC-2 FIR4\footnotemark[$\S$] \\
\hline
[CN]/[C\ss{15}N] & 230(80) & $290^{+160}_{-80}$ & 270(60) \\
$[$HCN]/[HC\ss{15}N] & 250(80)\footnotemark[$\|$] & $330^{+60}_{-50}$ & 270(50) \\
$[$HNC]/[H\ss{15}NC] & 300(100)\footnotemark[$\|$] & $225^{+75}_{-45}$ & 290(50) \\
\hline
\end{tabular}}\label{15Nratio}%
\begin{tabnote} 
\footnotemark[$*$] The numbers in parentheses represent the errors in units of the last significant digits.\\
\footnotemark[$\dagger$] This work\\
\footnotemark[$\ddagger$] \citet{daniel13}\\
\footnotemark[$\S$] \citet{kahane18}\\
\footnotemark[$\|$] The column density of the normal species are derived from the \ss{13}C species, where the \ss{12}C/\ss{13}C ratio is assumed to be 60.\\
\end{tabnote}
\end{table}
\clearpage
\begin{table}
\tbl{Column density ratios of organic species relative to CH$_3$OH\footnotemark[$*$]}{
\begin{tabular}{lcccc}
\hline
\multicolumn{1}{c}{Species} & L1527 & TMC-1 (CP\footnotemark[$\dagger$]) & TMC-1 (MP\footnotemark[$\ddagger$]) & IRAS16293-2422 \\
\hline
H$_2$CO&7(3)&--&--&1.1(8)\\
CH$_3$CHO&0.08(5)&0.046(16)\footnotemark[$\S$]&0.09(4)\footnotemark[$\S$]&0.06(3)\\
CH$_2$CO&0.31(19)&0.20(7)&0.09(4)&--\\
c-H$_2$C$_3$O&0.020(10)&0.011(2)&--&--\\
HCOOCH$_3$&$<0.42\;(3\sigma)$&--&0.026(11)\footnotemark[$\S$]&1.2(4)\\
(CH$_3$)$_2$O&$<0.22\;(3\sigma)$&$<0.13$\footnotemark[$\S$]&0.031(17)\footnotemark[$\S$]&11(6)\\
\hline
\end{tabular}}\label{COMs}%
\begin{tabnote} 
\footnotemark[$*$] The numbers in parentheses represent the errors in units of the last significant digits. The column densities are taken from \citet{soma18} for TMC-1. The beam-averaged column densities of CH$_3$OH and other COMs (CH$_3$CHO, HCOOCH$_3$, and (CH$_3$)$_2$O) in IRAS16293-2422 are evaluated by the least-squares fit of the observation data reported by \citet{caux11} and \citet{cazaux03}, respectively, under the assumption of the LTE condition.\\
\footnotemark[$\dagger$] Cyanopolyyne Peak\\
\footnotemark[$\ddagger$] Methanol Peak\\
\footnotemark[$\S$] The column density is the sum of two velocity components.\\
\end{tabnote}
\end{table}
\clearpage
\begin{table}
\tbl{Comparison of rest frequencies of the CCD $N=1$--0 transition between our observation and calculation\footnotemark[$*$]}{
\begin{tabular}{ccc}
\hline
Transition & \multicolumn{2}{c}{Rest frequency (GHz)} \\
\cline{2-3}
&Observation\footnotemark[$\dagger$]&Calculation\footnotemark[$\ddagger$]\\
\hline
$N=$1--0, $J=$3/2--1/2, $F=$3/2--3/2&72.101780(10)&72.1017155\\
$N=$1--0, $J=$3/2--1/2, $F=$5/2--3/2&72.1077085(3)&72.1077000\\
$N=$1--0, $J=$3/2--1/2, $F=$1/2--1/2&72.1090173(7)&72.1091138\\
$N=$1--0, $J=$3/2--1/2, $F=$3/2--1/2&72.112277(5)&72.1123994\\
$N=$1--0, $J=$1/2--1/2, $F=$3/2--3/2&72.187710(6)&72.1877041\\
$N=$1--0, $J=$1/2--1/2, $F=$1/2--3/2&72.189729(15)&72.1895050\\
$N=$1--0, $J=$1/2--1/2, $F=$3/2--1/2&72.198193(12)&72.1983880\\
\hline
\end{tabular}}\label{tab_CCD}%
\begin{tabnote}
\footnotemark[$*$] The numbers in parentheses represent the errors in units of last significant digits.\\
\footnotemark[$\dagger$] $V_\mathrm{LSR}$ of $5.85\,\mathrm{km\,s^{-1}}$ is assumed.\\
\footnotemark[$\ddagger$] Taken from CDMS.\\
\end{tabnote}
\end{table}

\clearpage
\clearpage
\begin{figure}[!htb]
\centering
\includegraphics[height=.99\textheight]{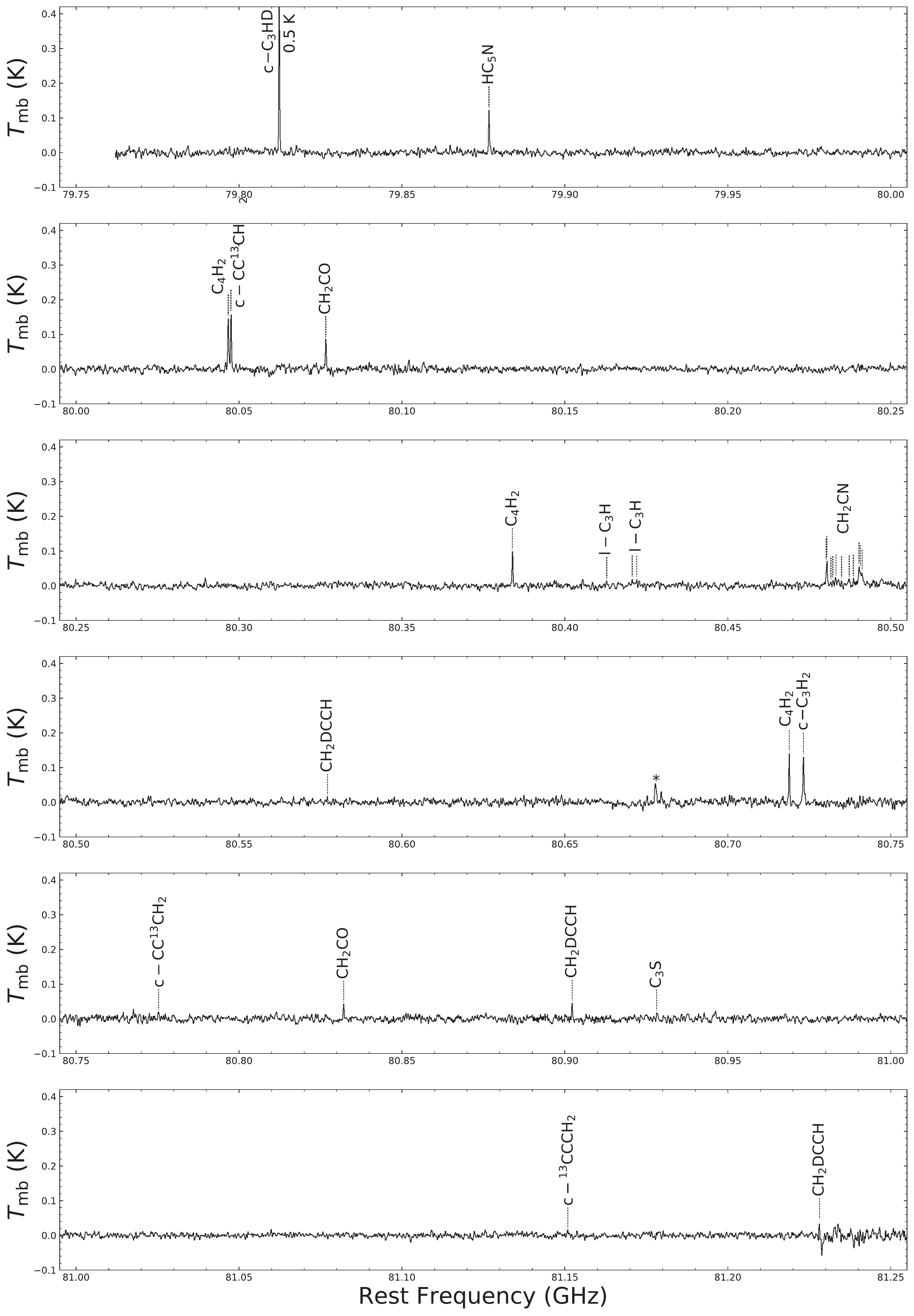}
\caption{Spectrum of L1527 in the 3 mm band. Spurious lines are indicated with asterisks.}\label{spect_3mm}
\end{figure}
\clearpage
\addtocounter{figure}{-1}
\begin{figure}[!htb]
 \centering
 \includegraphics[height=.99\textheight]{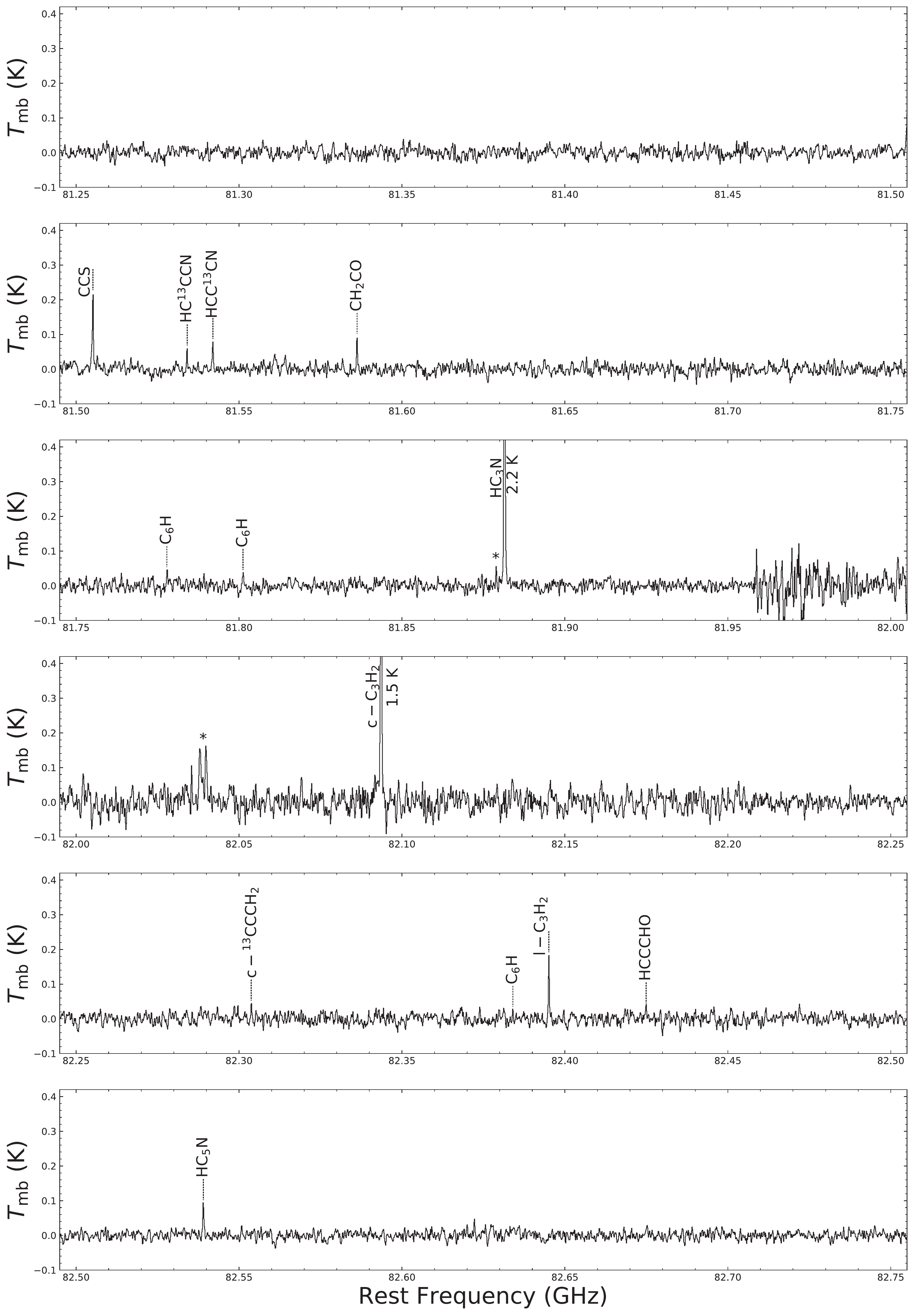}
\caption{(Continued.)}
\end{figure}
\clearpage
\addtocounter{figure}{-1}
\begin{figure}[!htb]
\centering
\includegraphics[height=.99\textheight]{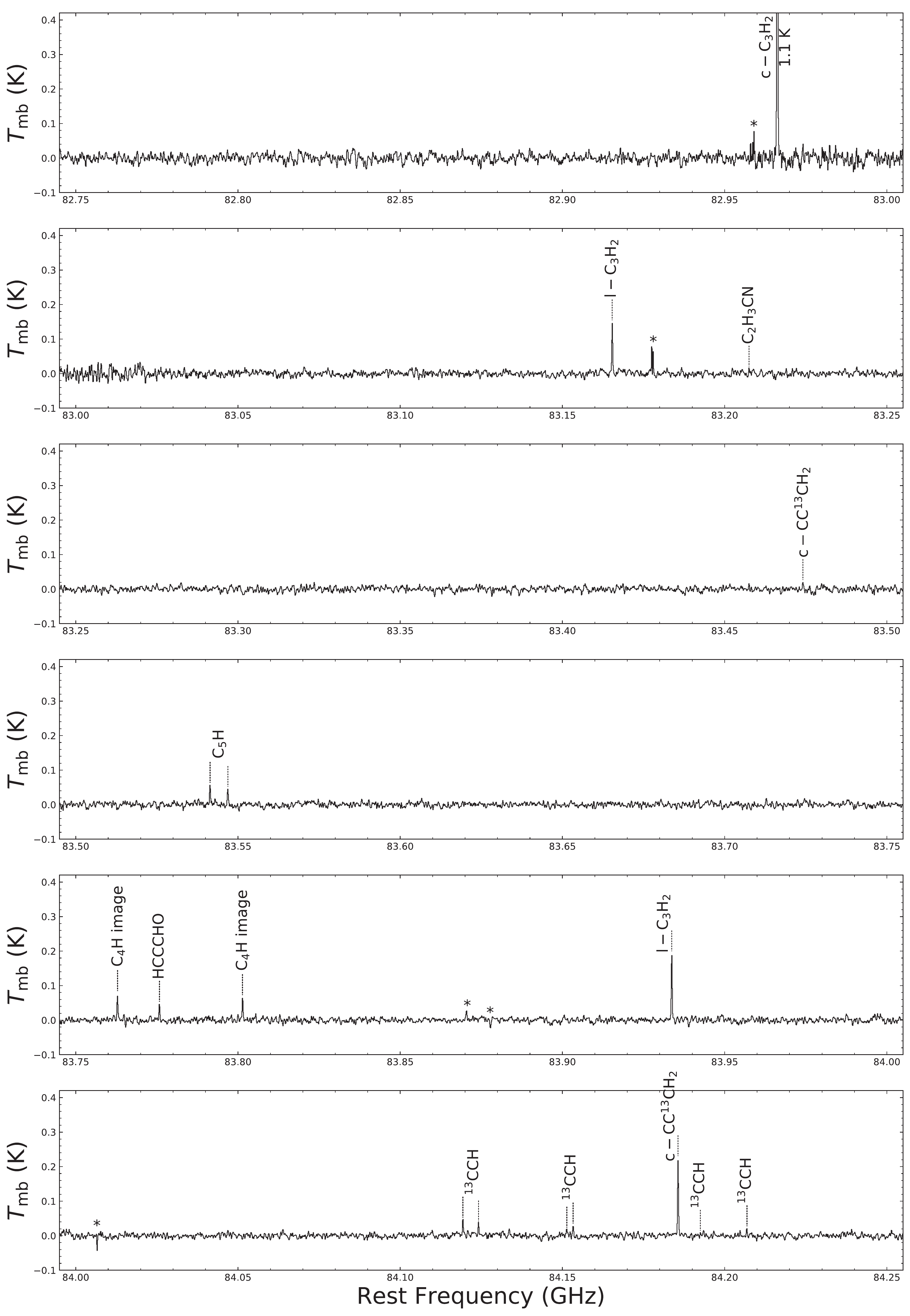}
\caption{(Continued.)}
\end{figure}
\clearpage
\addtocounter{figure}{-1}
\begin{figure}[!htb]
\centering
\includegraphics[height=.99\textheight]{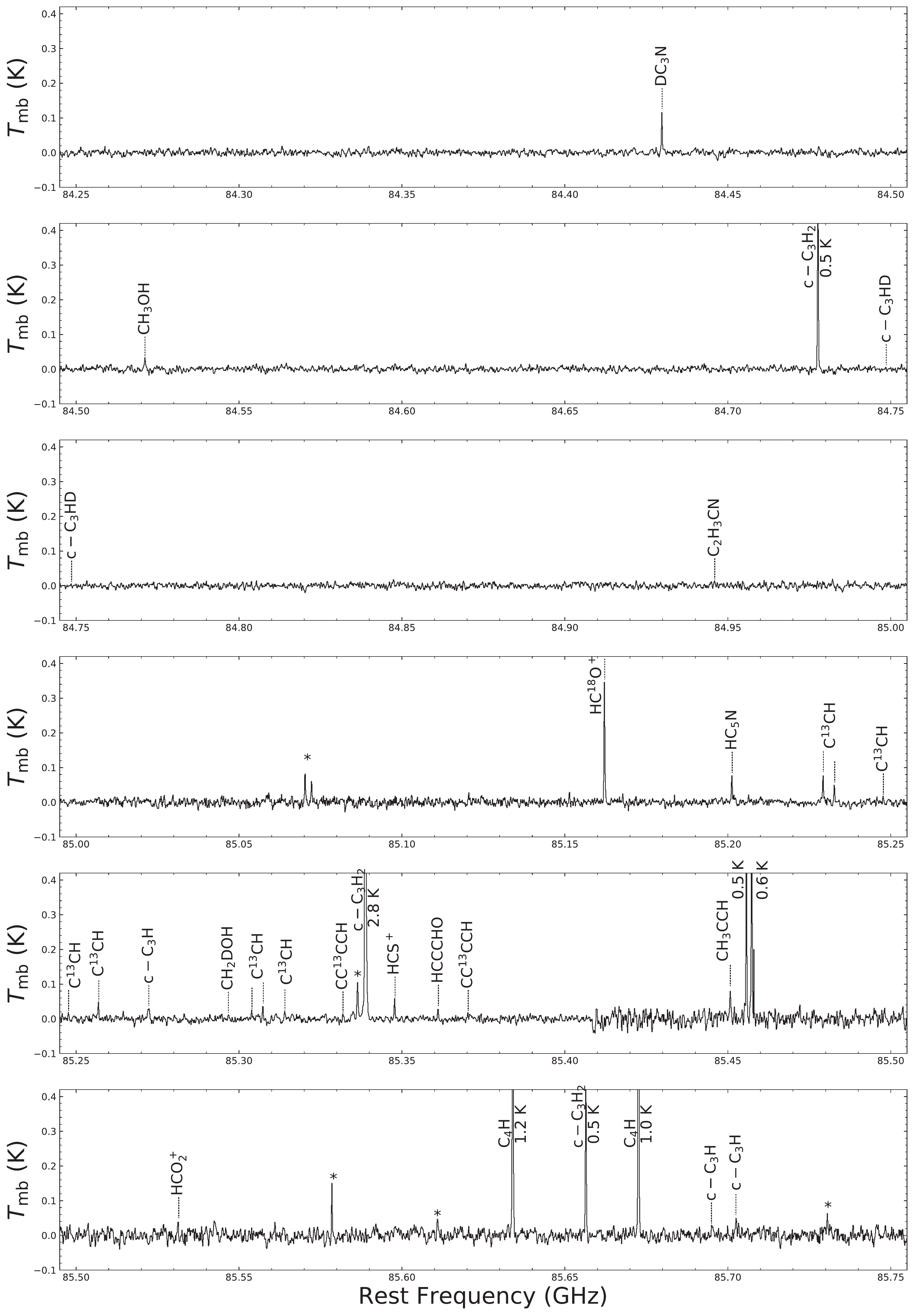}
\caption{(Continued.)}
\end{figure}
\clearpage
\addtocounter{figure}{-1}
\begin{figure}[!htb]
\centering
\includegraphics[height=.99\textheight]{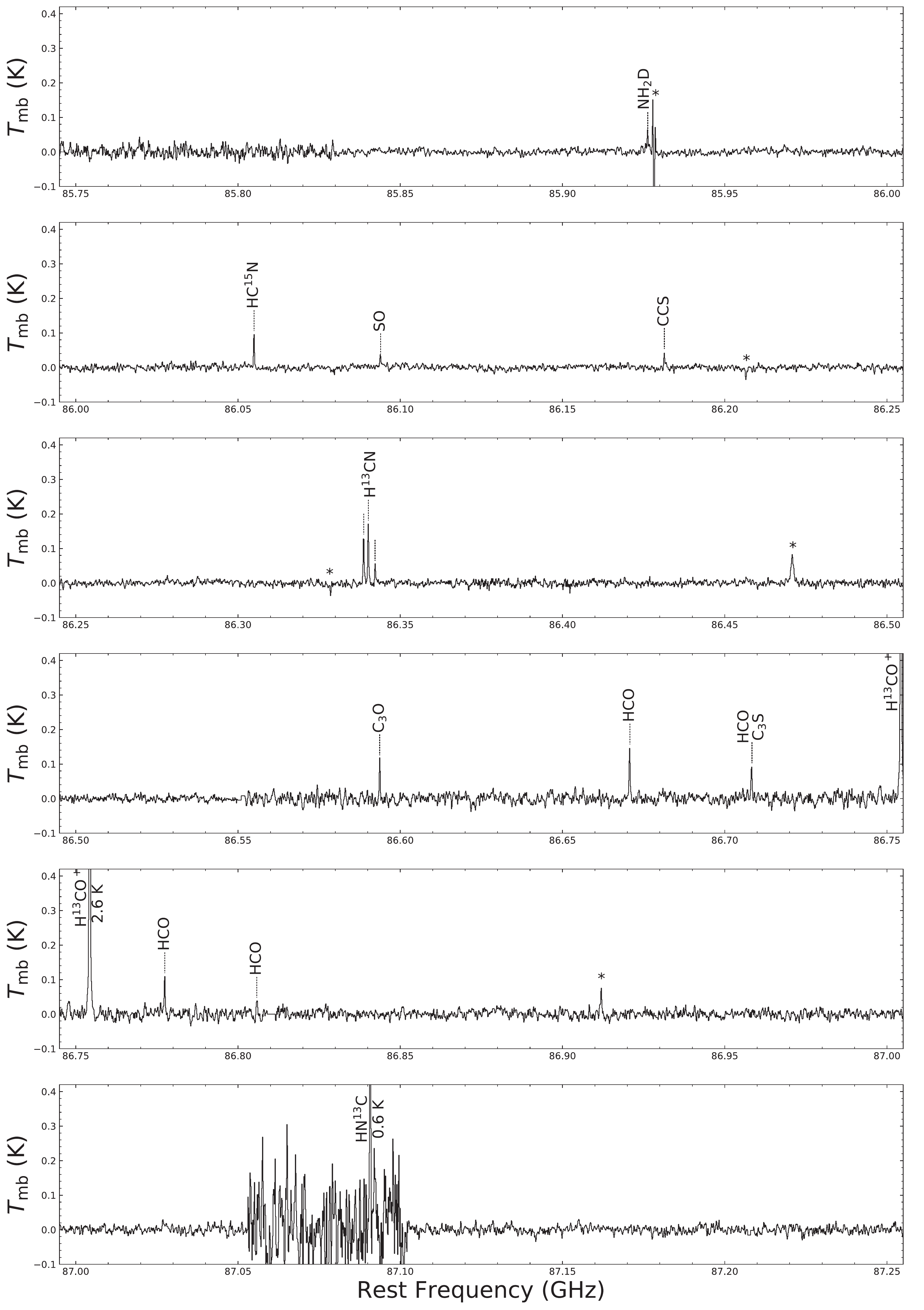}
\caption{(Continued.)}
\end{figure}
\clearpage
\addtocounter{figure}{-1}
\begin{figure}[!htb]
\centering
\includegraphics[height=.99\textheight]{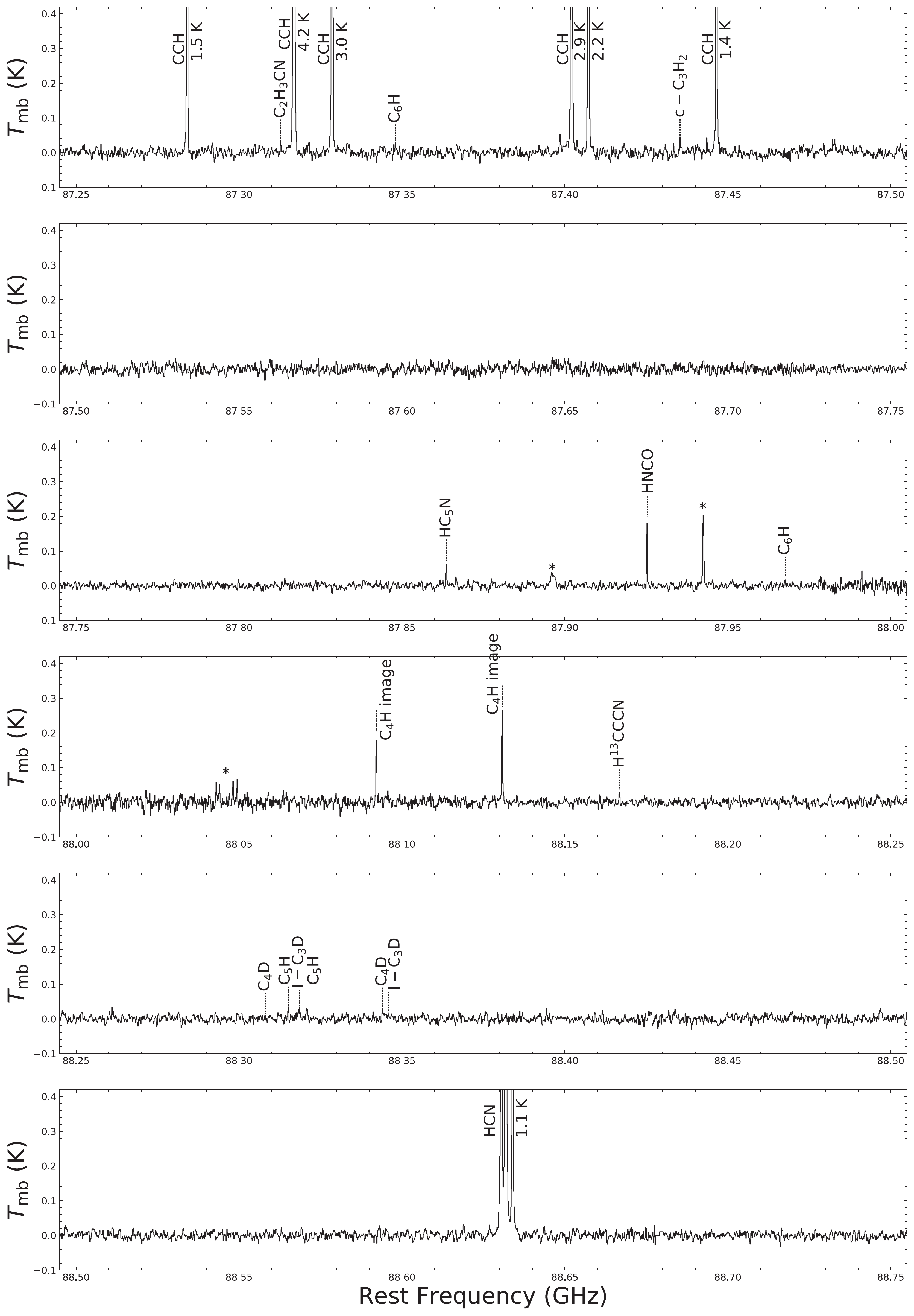}
\caption{(Continued.)}
\end{figure}
\clearpage
\addtocounter{figure}{-1}
\begin{figure}[!htb]
\centering
\includegraphics[height=.99\textheight]{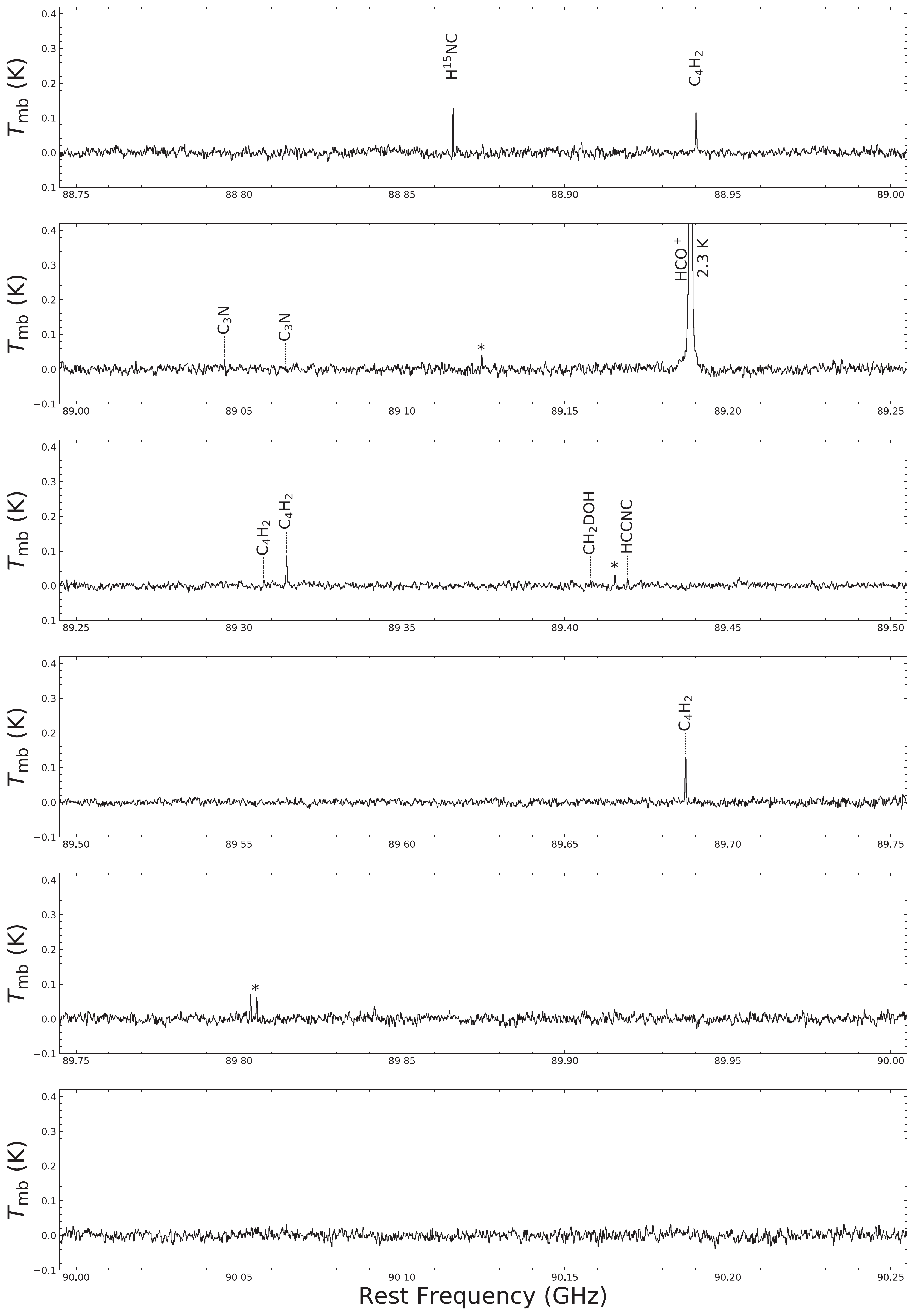}
\caption{(Continued.)}
\end{figure}
\clearpage
\addtocounter{figure}{-1}
\begin{figure}[!htb]
\centering
\includegraphics[height=.99\textheight]{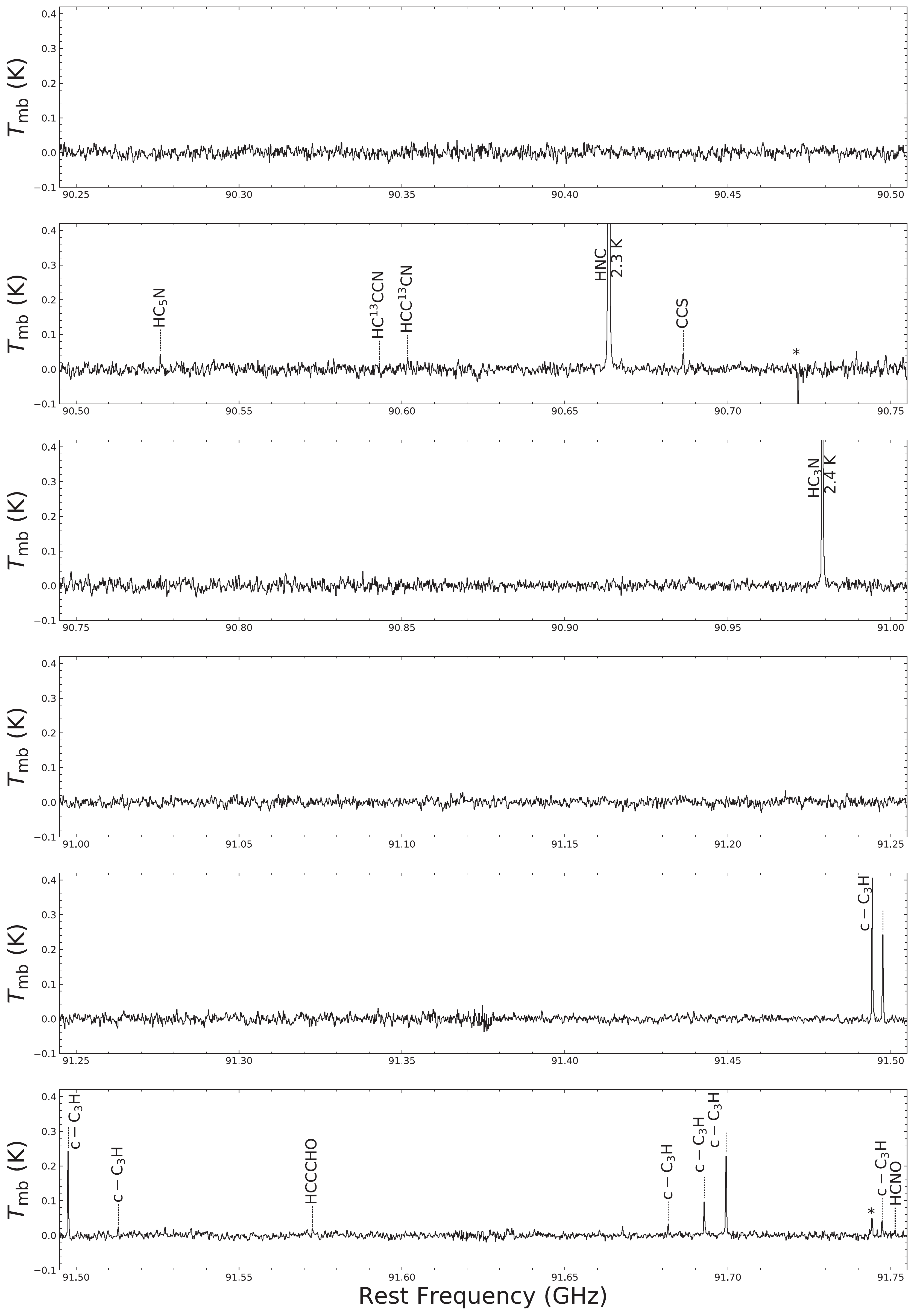}
\caption{(Continued.)}
\end{figure}
\clearpage
\addtocounter{figure}{-1}
\begin{figure}[!htb]
\centering
\includegraphics[height=.99\textheight]{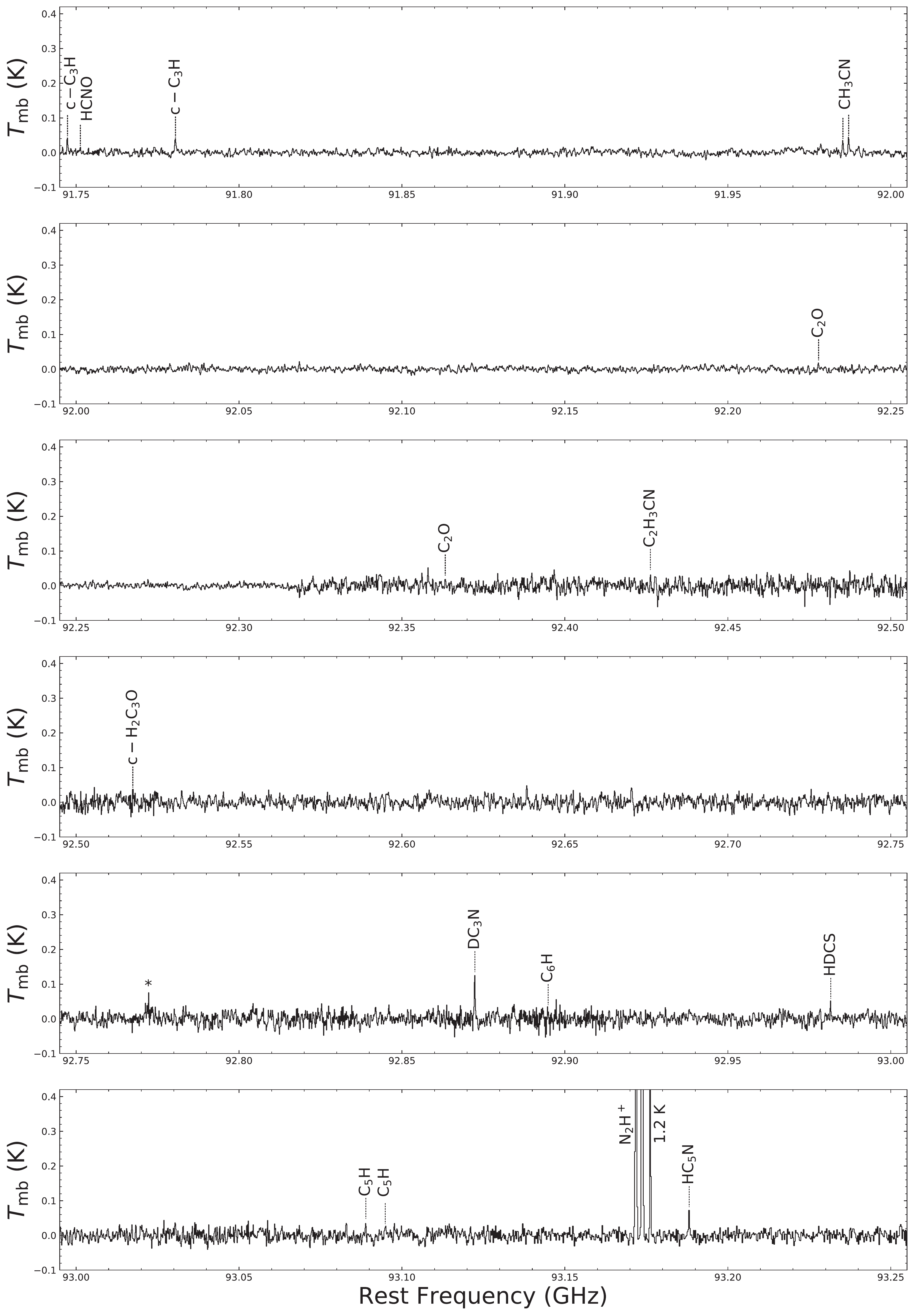}
\caption{(Continued.)}
\end{figure}
\clearpage
\addtocounter{figure}{-1}
\begin{figure}[!htb]
\centering
\includegraphics[height=.99\textheight]{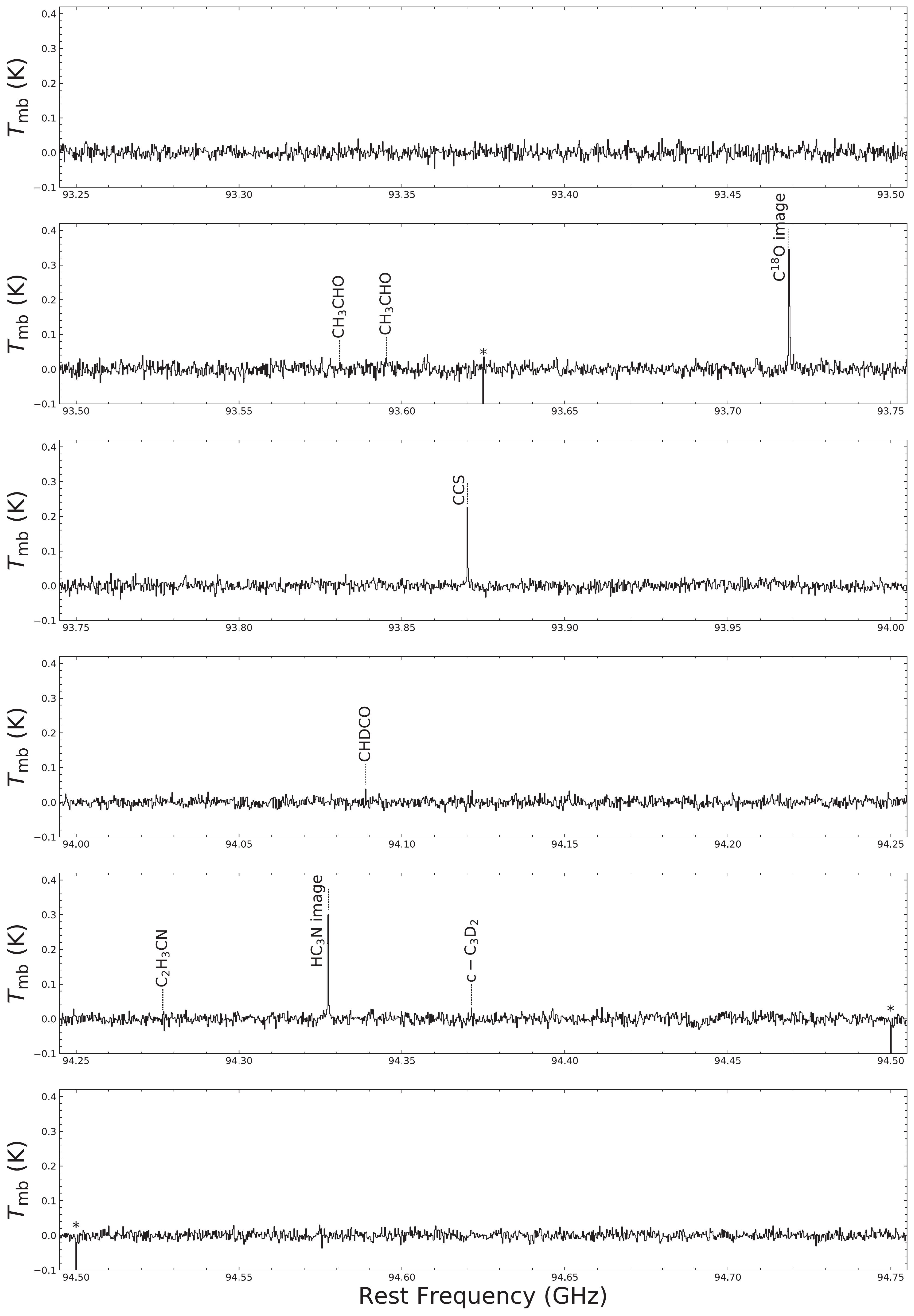}
\caption{(Continued.)}
\end{figure}
\clearpage
\addtocounter{figure}{-1}
\begin{figure}[!htb]
\centering
\includegraphics[height=.99\textheight]{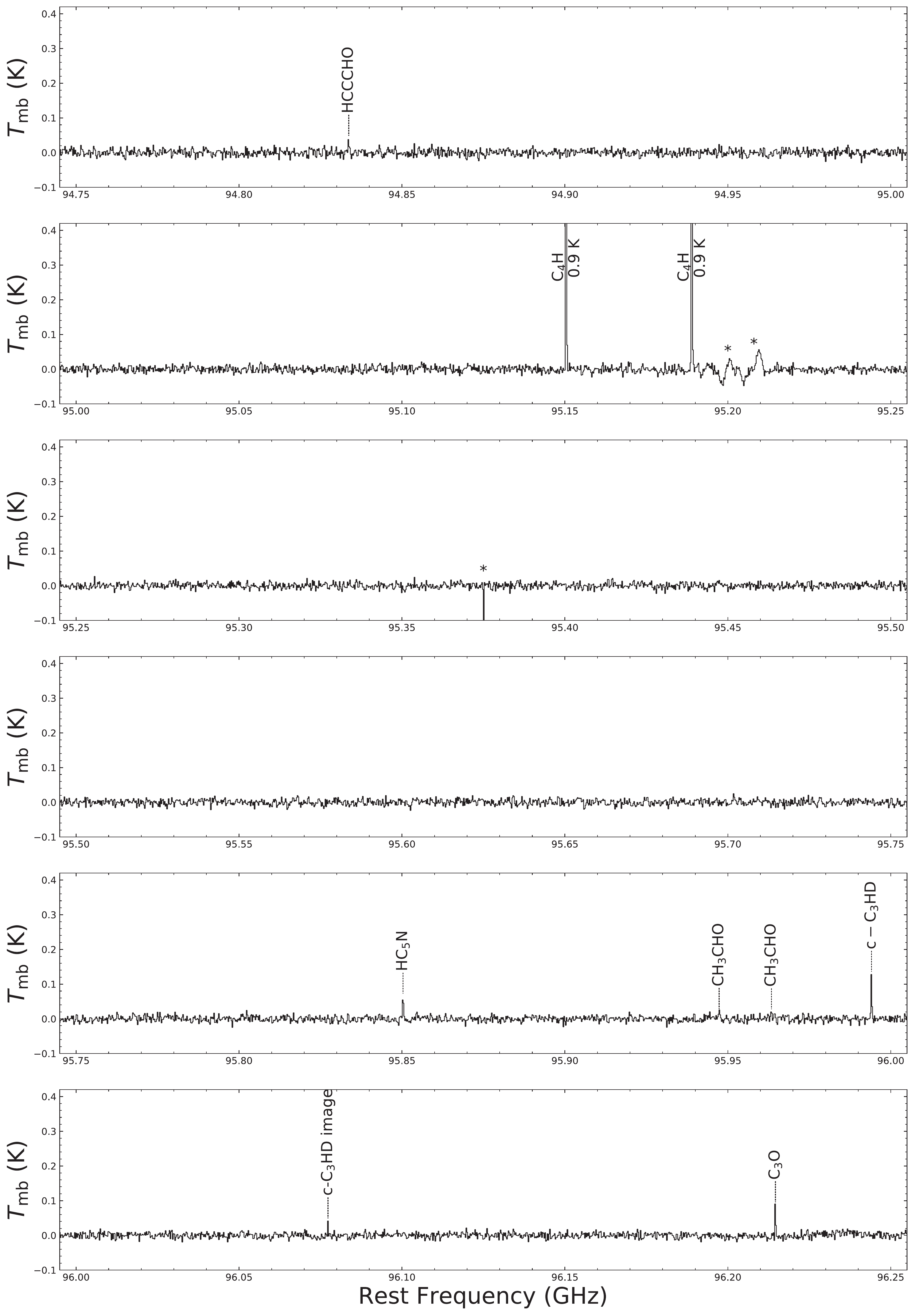}
\caption{(Continued.)}
\end{figure}
\clearpage
\addtocounter{figure}{-1}
\begin{figure}[!htb]
\centering
\includegraphics[height=.99\textheight]{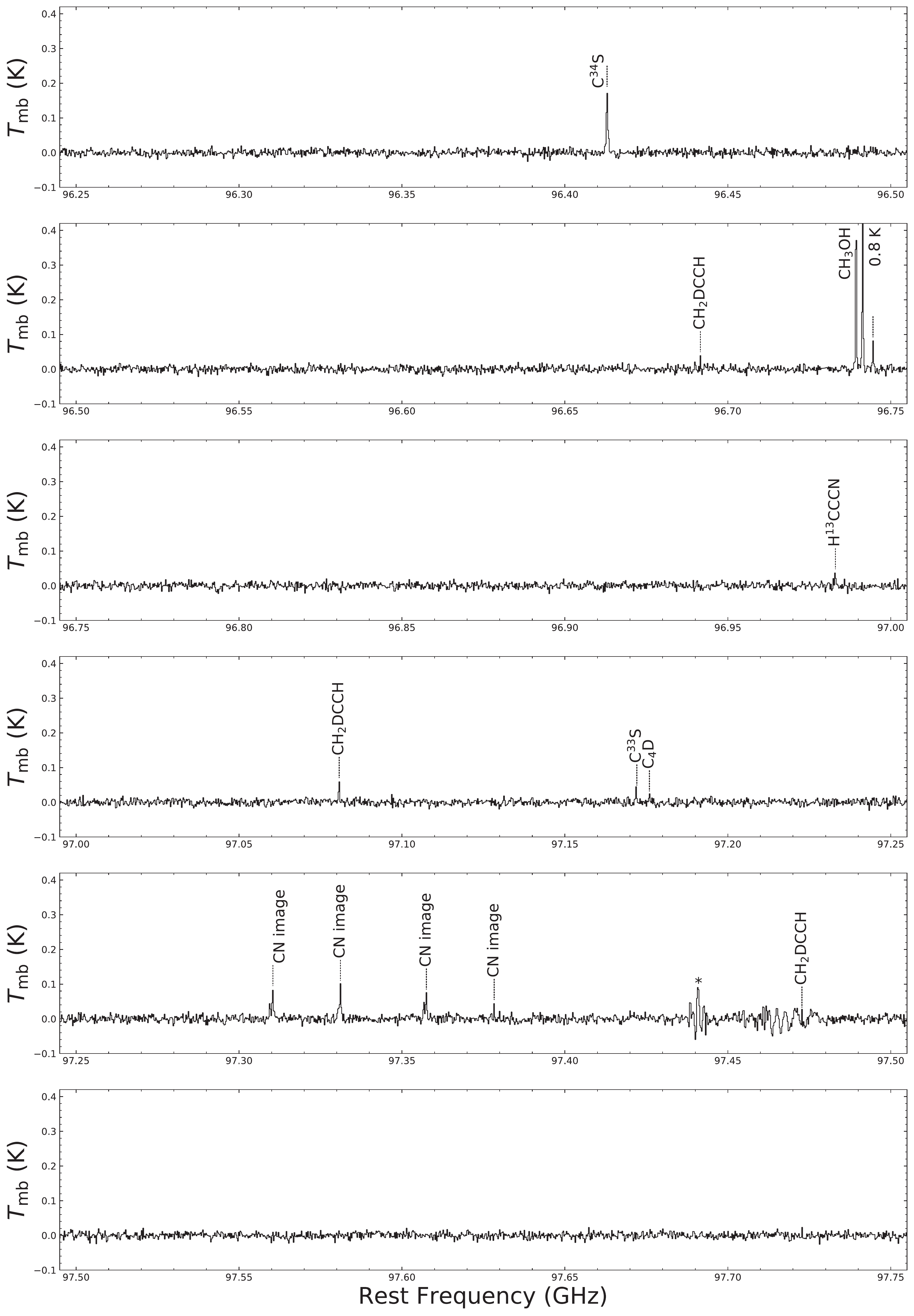}
\caption{(Continued.)}
\end{figure}
\clearpage
\addtocounter{figure}{-1}
\begin{figure}[!htb]
\centering
\includegraphics[height=.99\textheight]{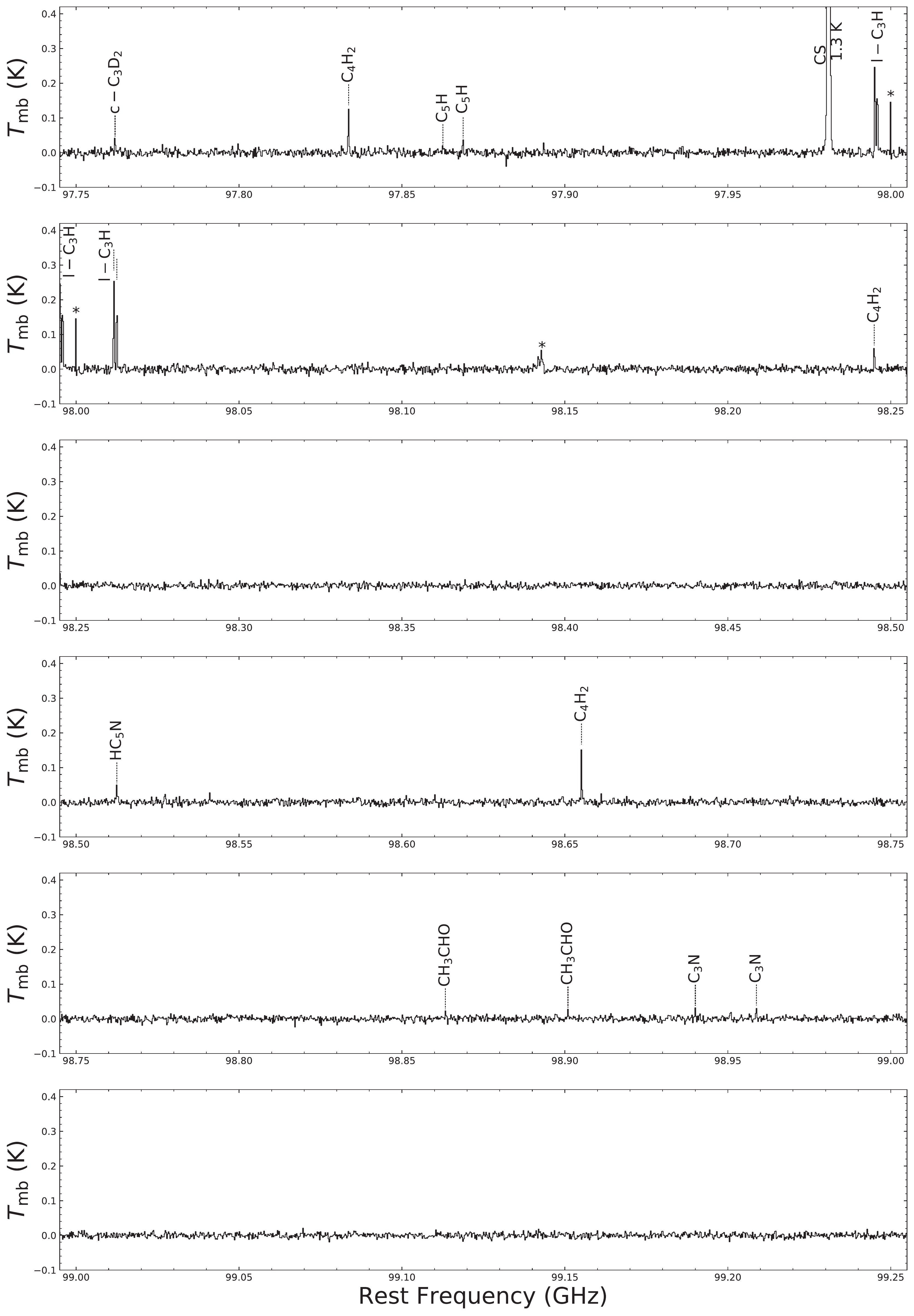}
\caption{(Continued.)}
\end{figure}
\clearpage
\addtocounter{figure}{-1}
\begin{figure}[!htb]
\centering
\includegraphics[height=.99\textheight]{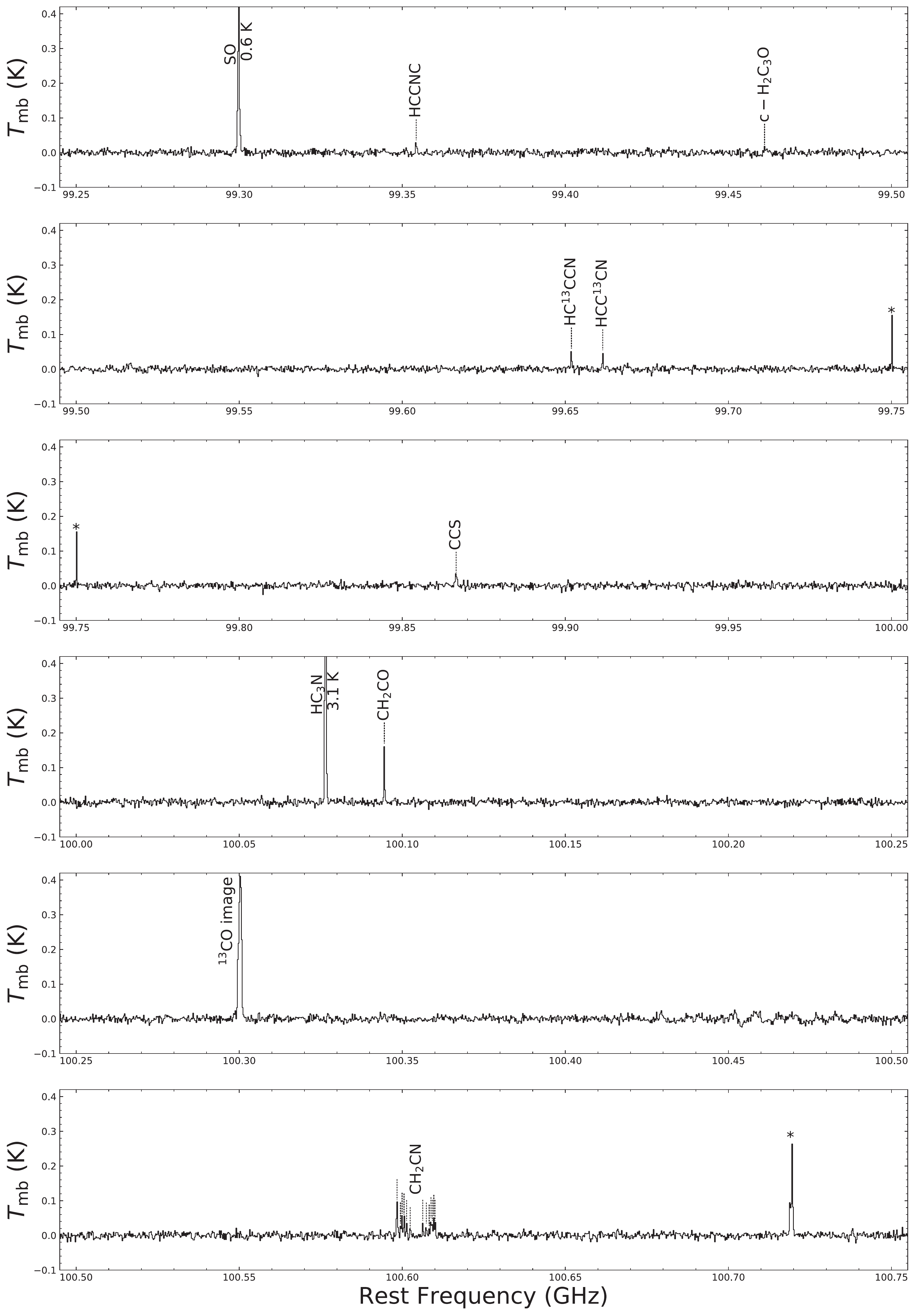}
\caption{(Continued.)}
\end{figure}
\clearpage
\addtocounter{figure}{-1}
\begin{figure}[!htb]
\centering
\includegraphics[height=.99\textheight]{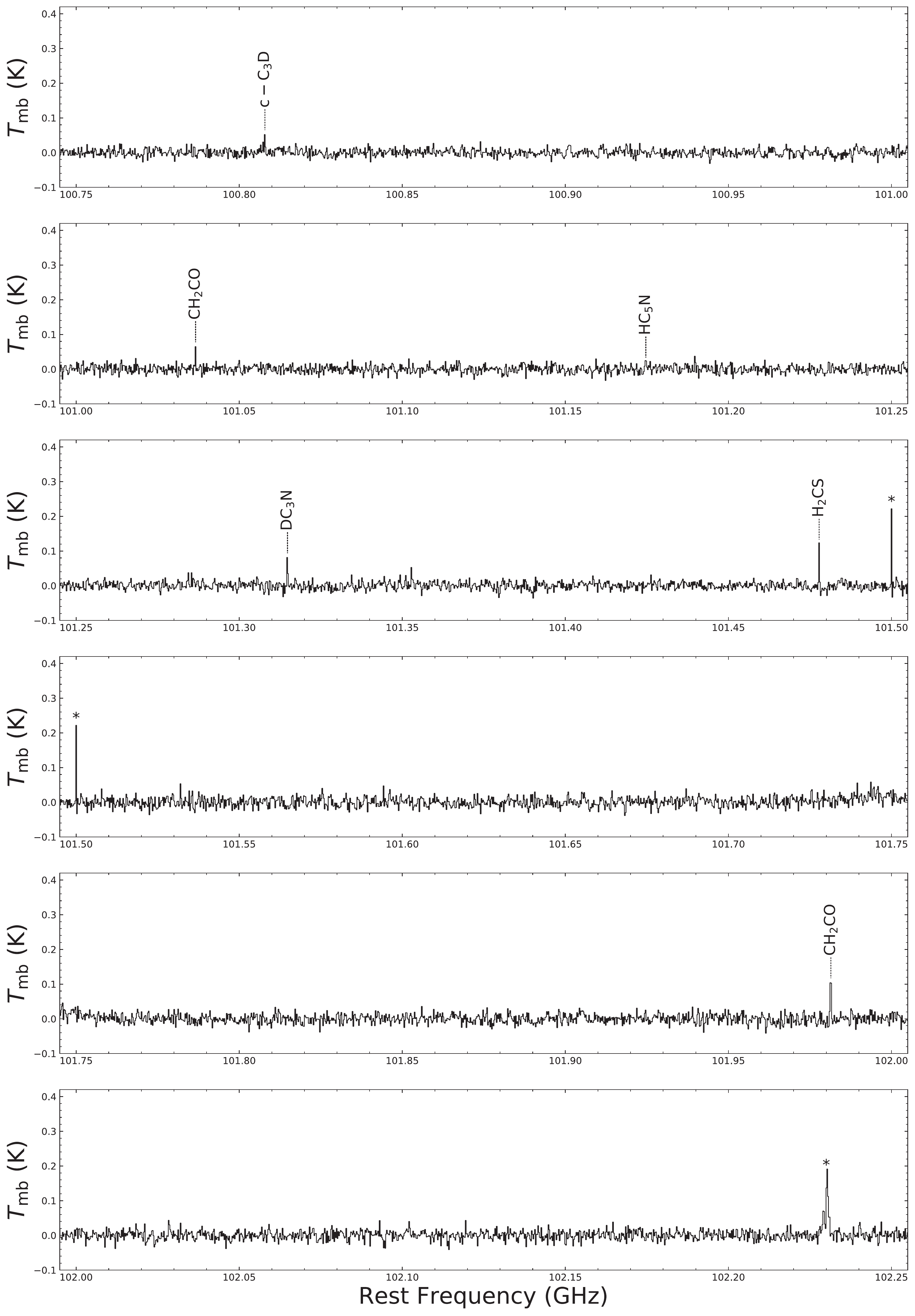}
\caption{(Continued.)}
\end{figure}
\clearpage
\addtocounter{figure}{-1}
\begin{figure}[!htb]
\centering
\includegraphics[height=.99\textheight]{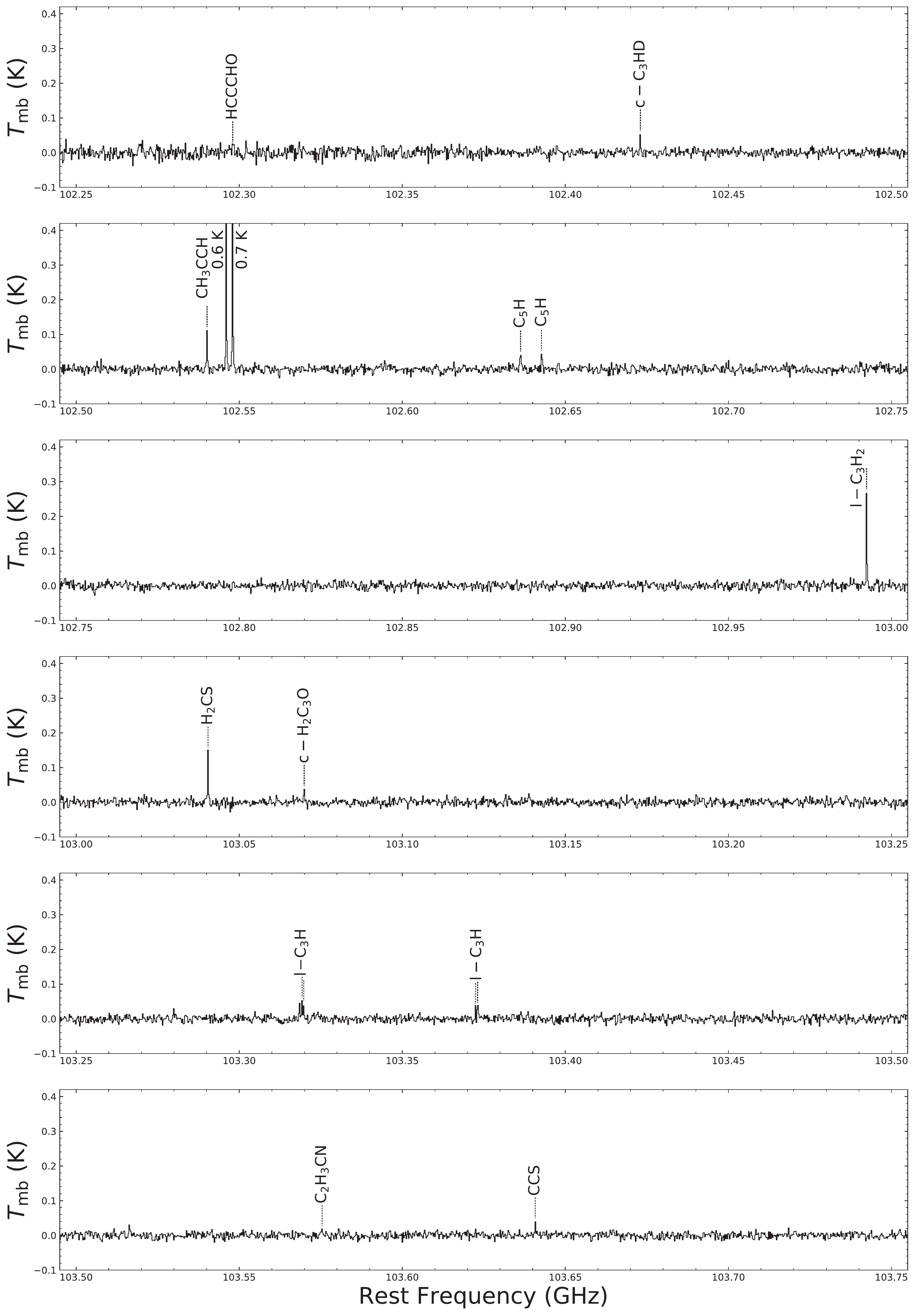}
\caption{(Continued.)}
\end{figure}
\clearpage
\addtocounter{figure}{-1}
\begin{figure}[!htb]
\centering
\includegraphics[height=.99\textheight]{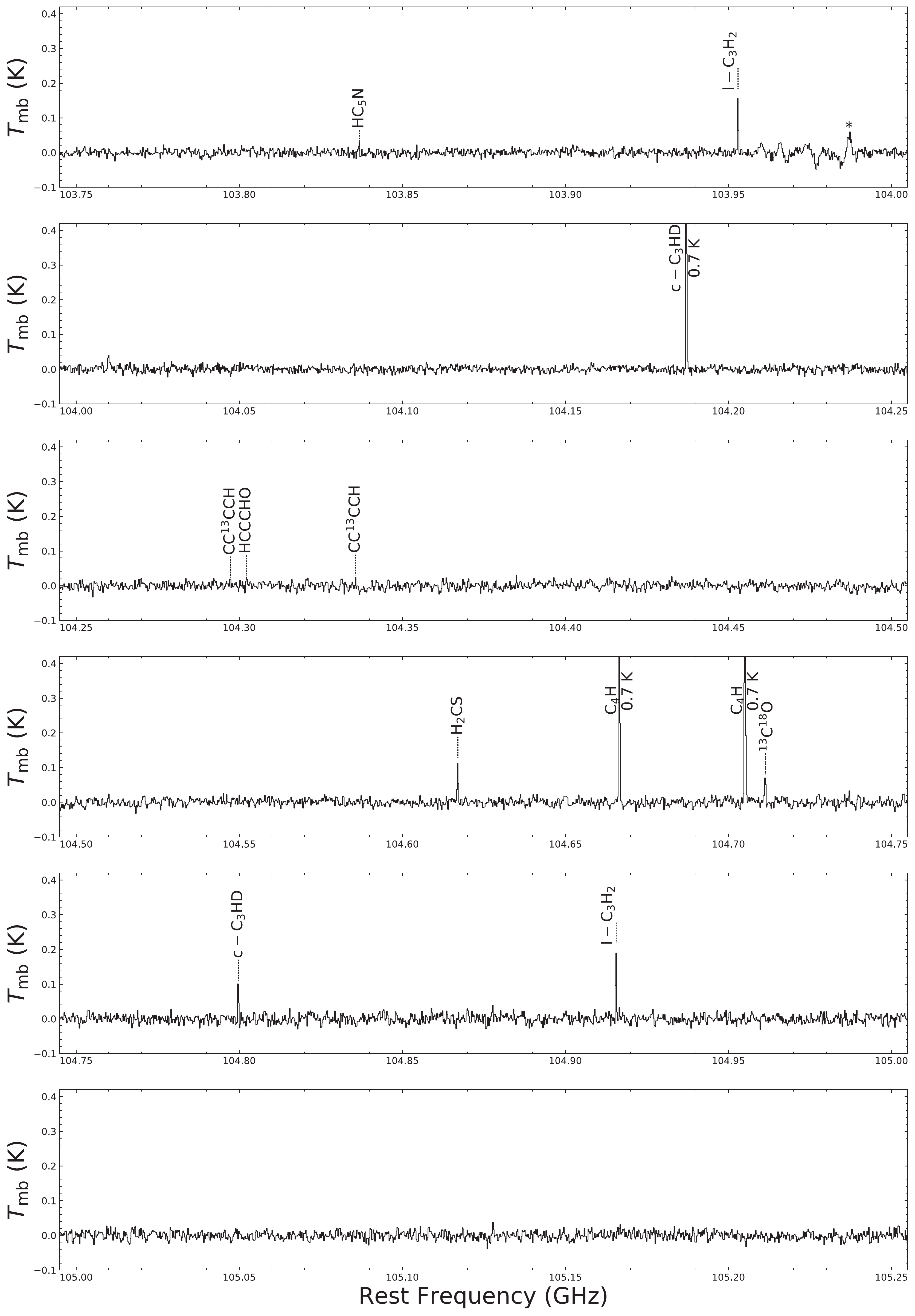}
\caption{(Continued.)}
\end{figure}
\clearpage
\addtocounter{figure}{-1}
\begin{figure}[!htb]
\centering
\includegraphics[height=.99\textheight]{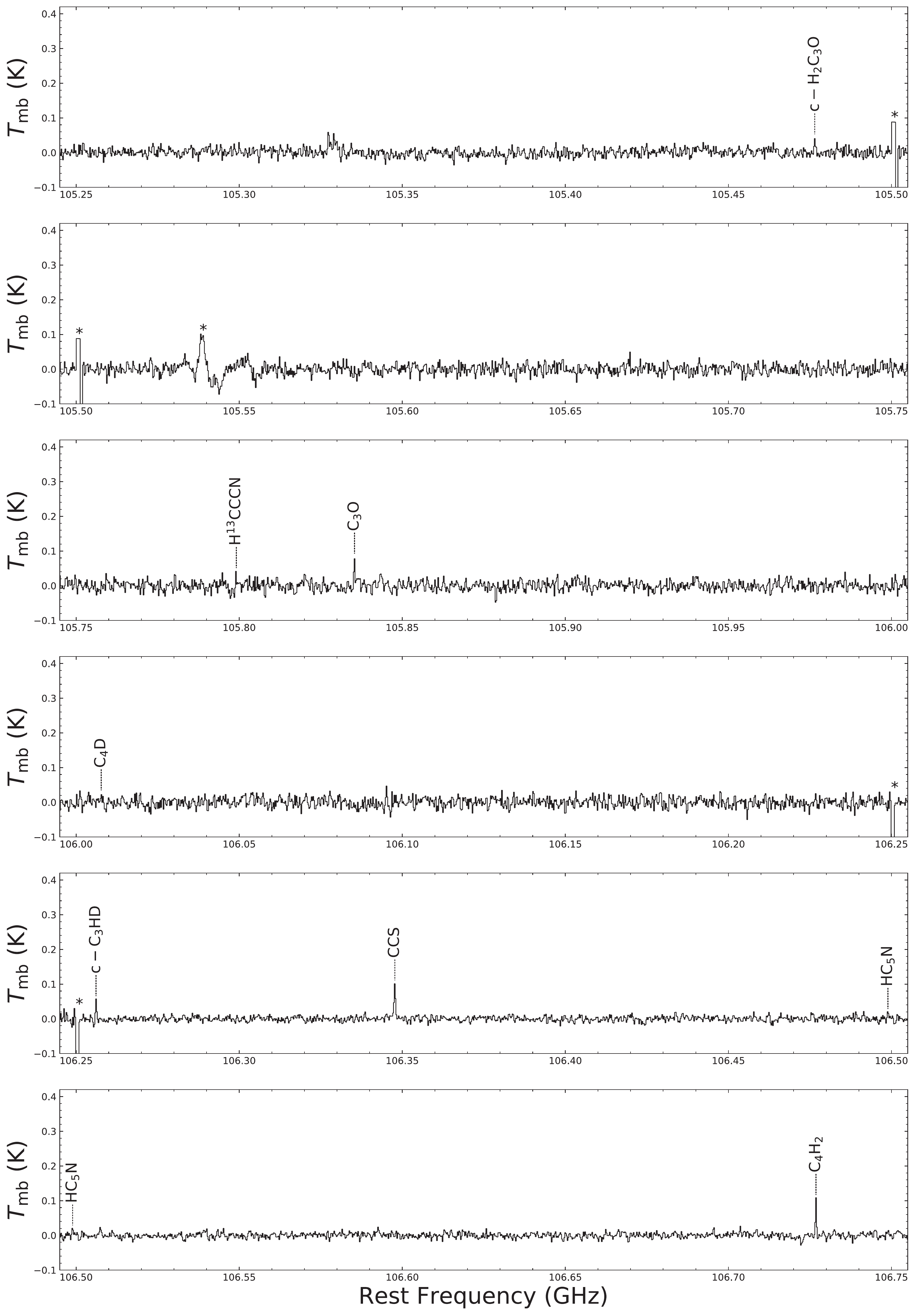}
\caption{(Continued.)}
\end{figure}
\clearpage
\addtocounter{figure}{-1}
\begin{figure}[!htb]
\centering
\includegraphics[height=.99\textheight]{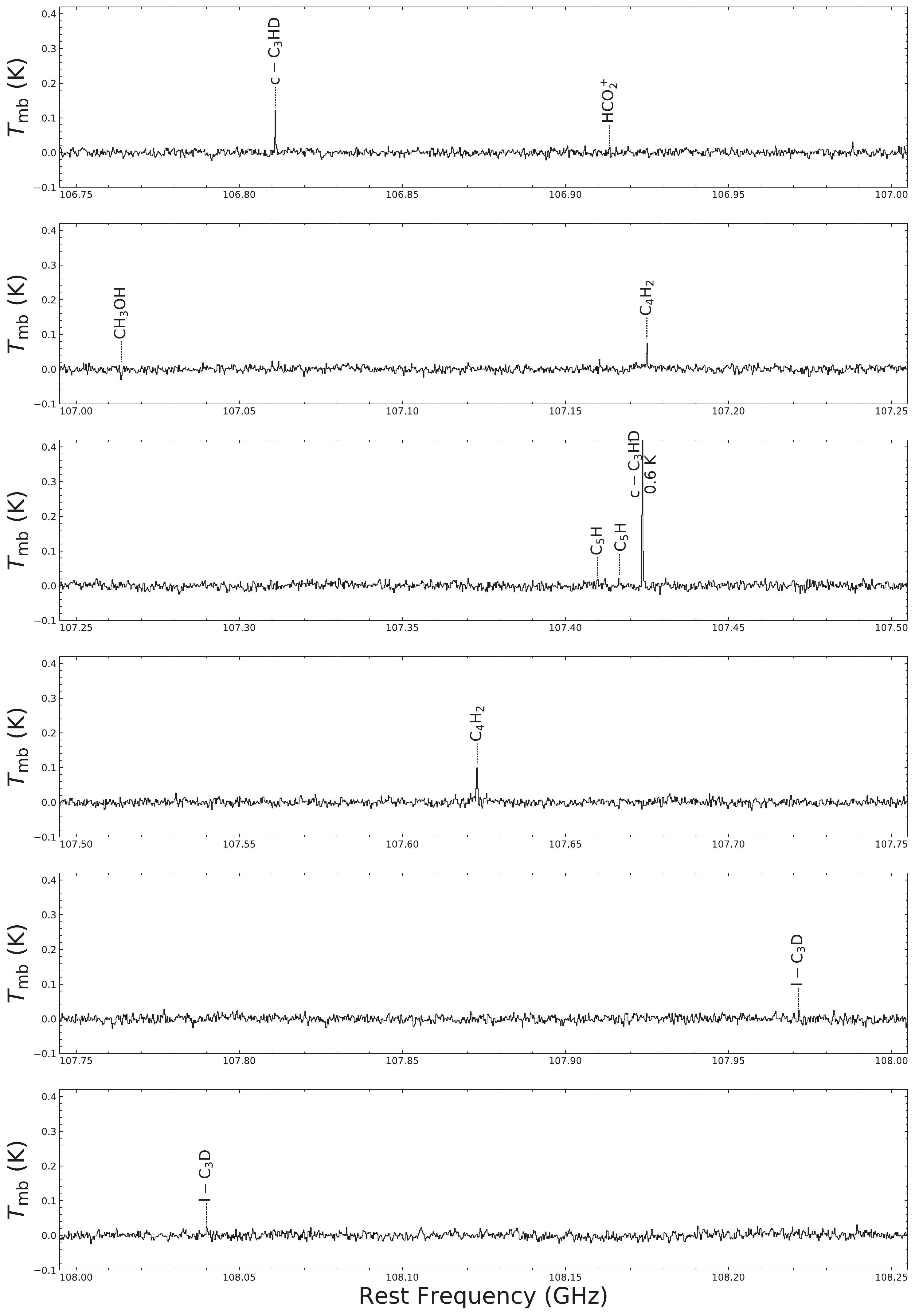}
\caption{(Continued.)}
\end{figure}
\clearpage
\addtocounter{figure}{-1}
\begin{figure}[!htb]
\centering
\includegraphics[height=.99\textheight]{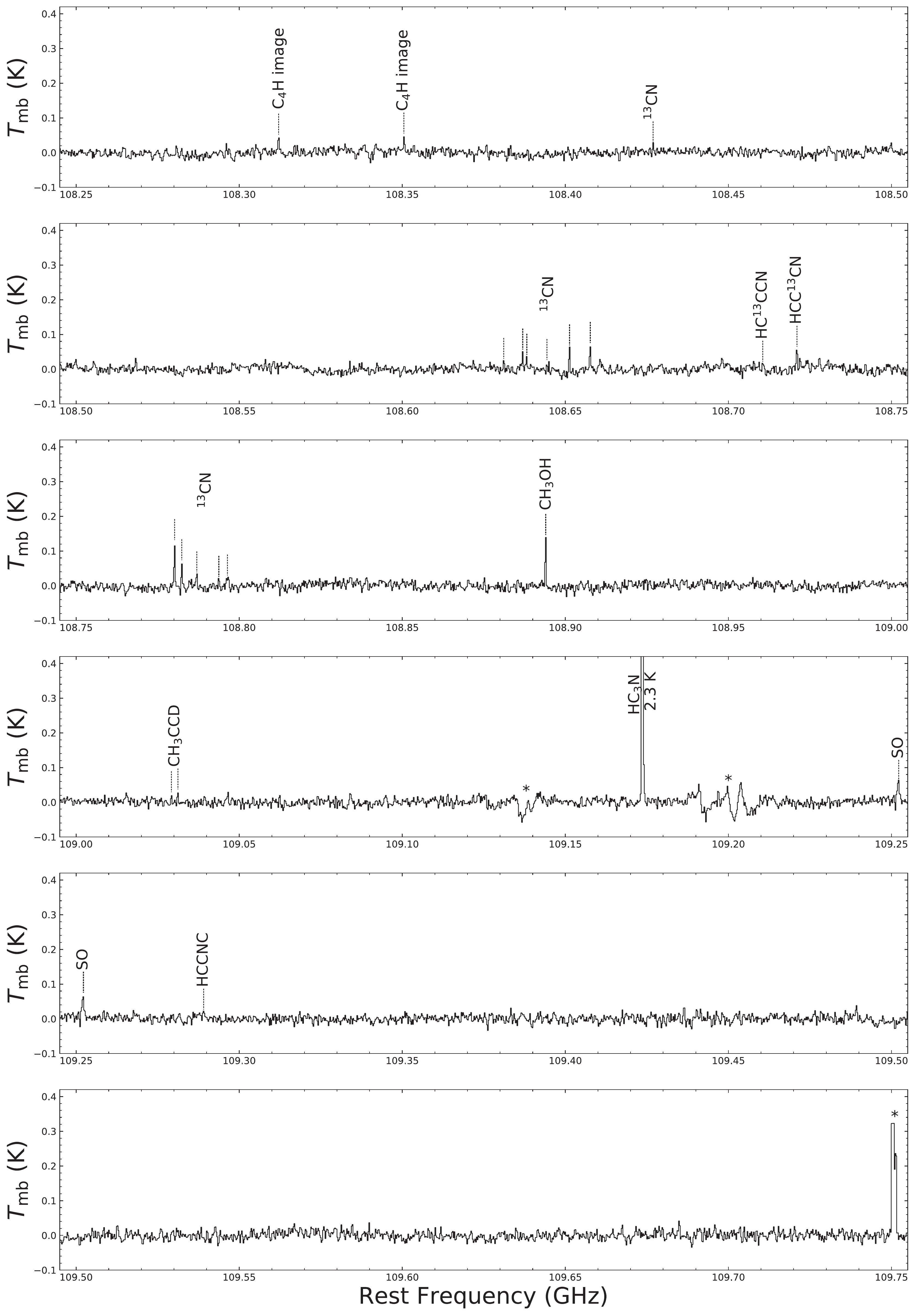}
\caption{(Continued.)}
\end{figure}
\clearpage
\addtocounter{figure}{-1}
\begin{figure}[!htb]
\centering
\includegraphics[height=.99\textheight]{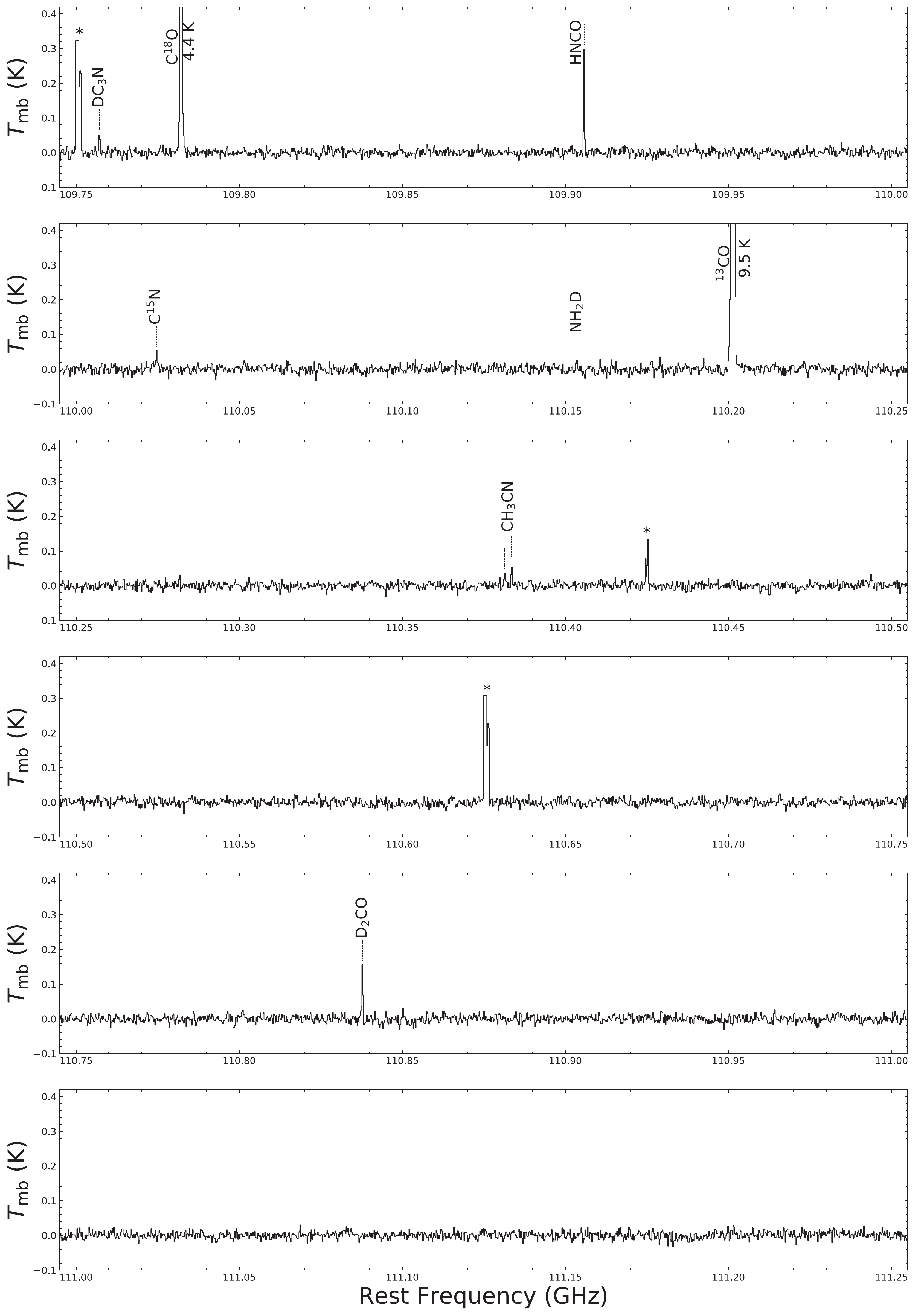}
\caption{(Continued.)}
\end{figure}
\clearpage
\addtocounter{figure}{-1}
\begin{figure}[!htb]
\centering
\includegraphics[height=.99\textheight]{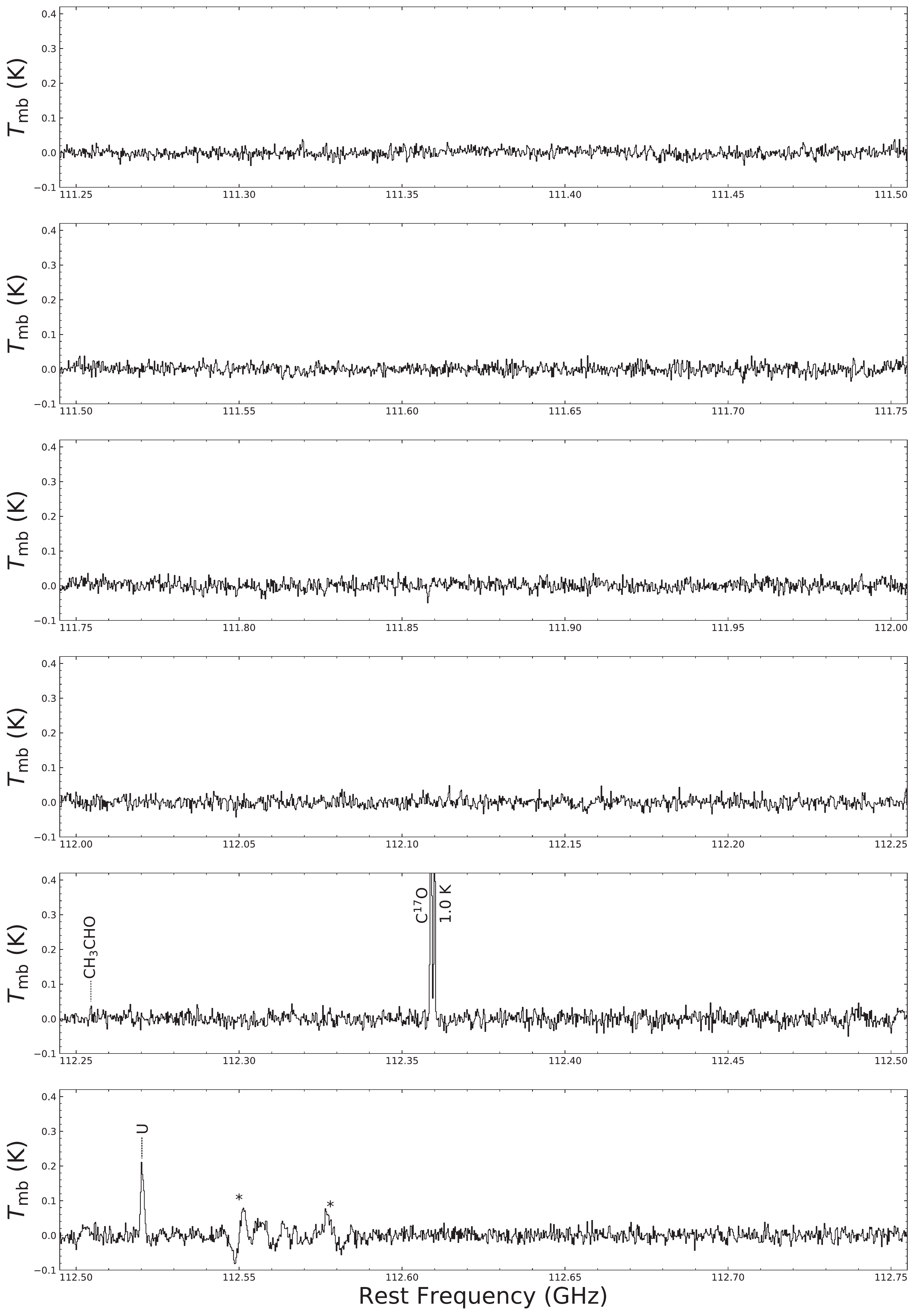}
\caption{(Continued.)}
\end{figure}
\clearpage
\addtocounter{figure}{-1}
\begin{figure}[!htb]
\centering
\includegraphics[height=.99\textheight]{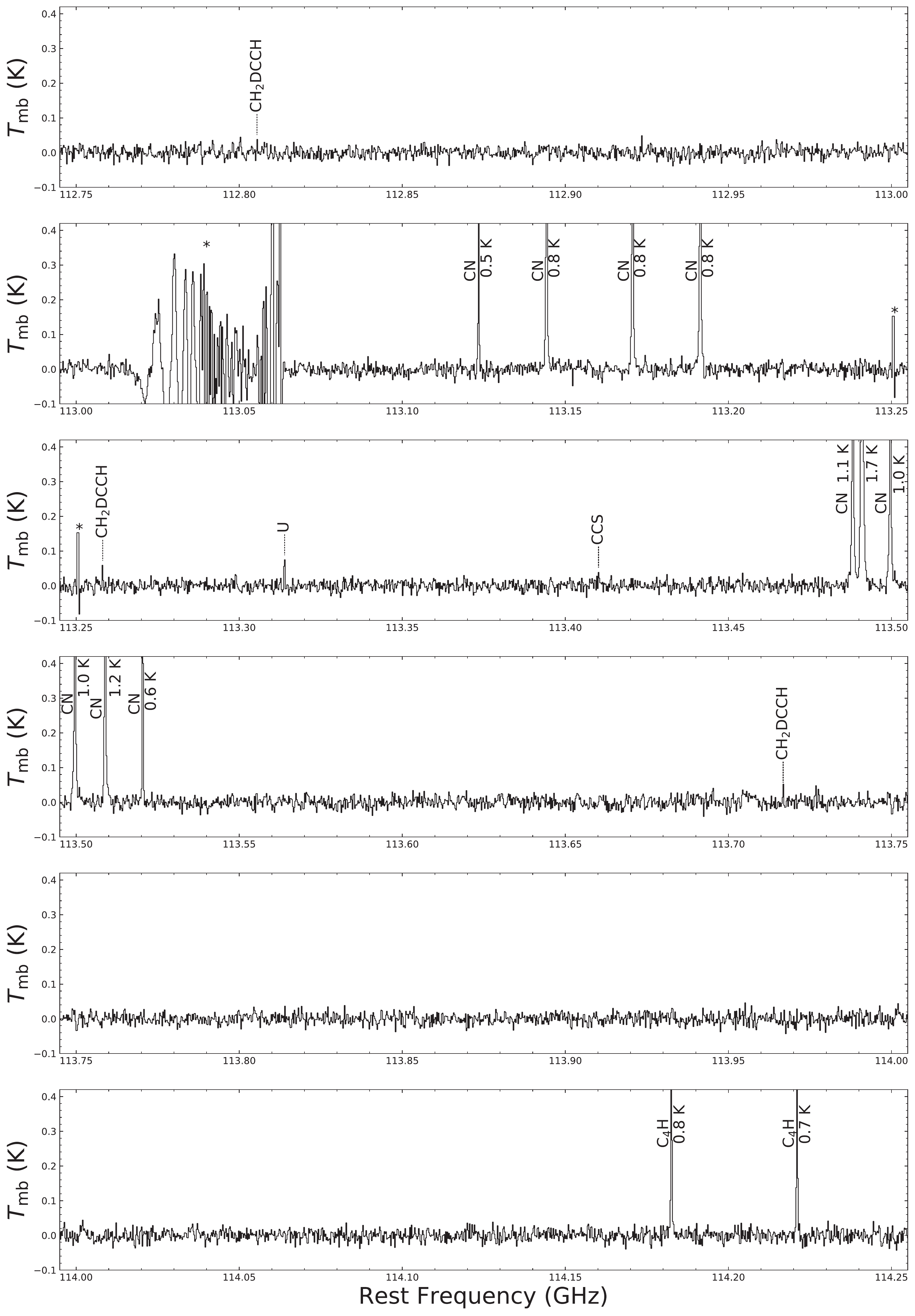}
\caption{(Continued.)}
\end{figure}
\clearpage
\addtocounter{figure}{-1}
\begin{figure}[!htb]
\centering
\includegraphics[height=.99\textheight]{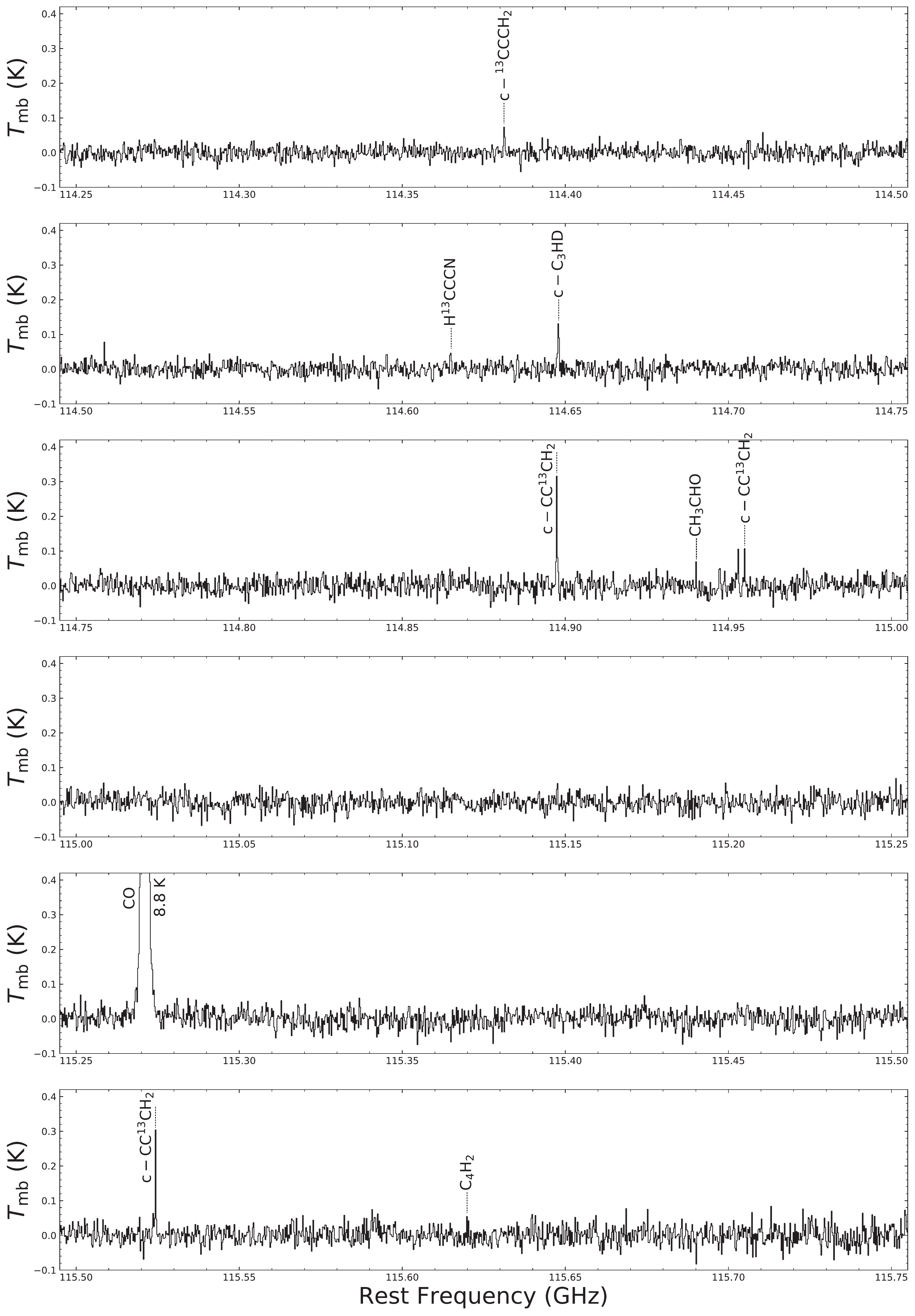}
\caption{(Continued.)}
\end{figure}
\clearpage
\addtocounter{figure}{-1}
\begin{figure}[!htb]
\centering
\includegraphics[height=.99\textheight]{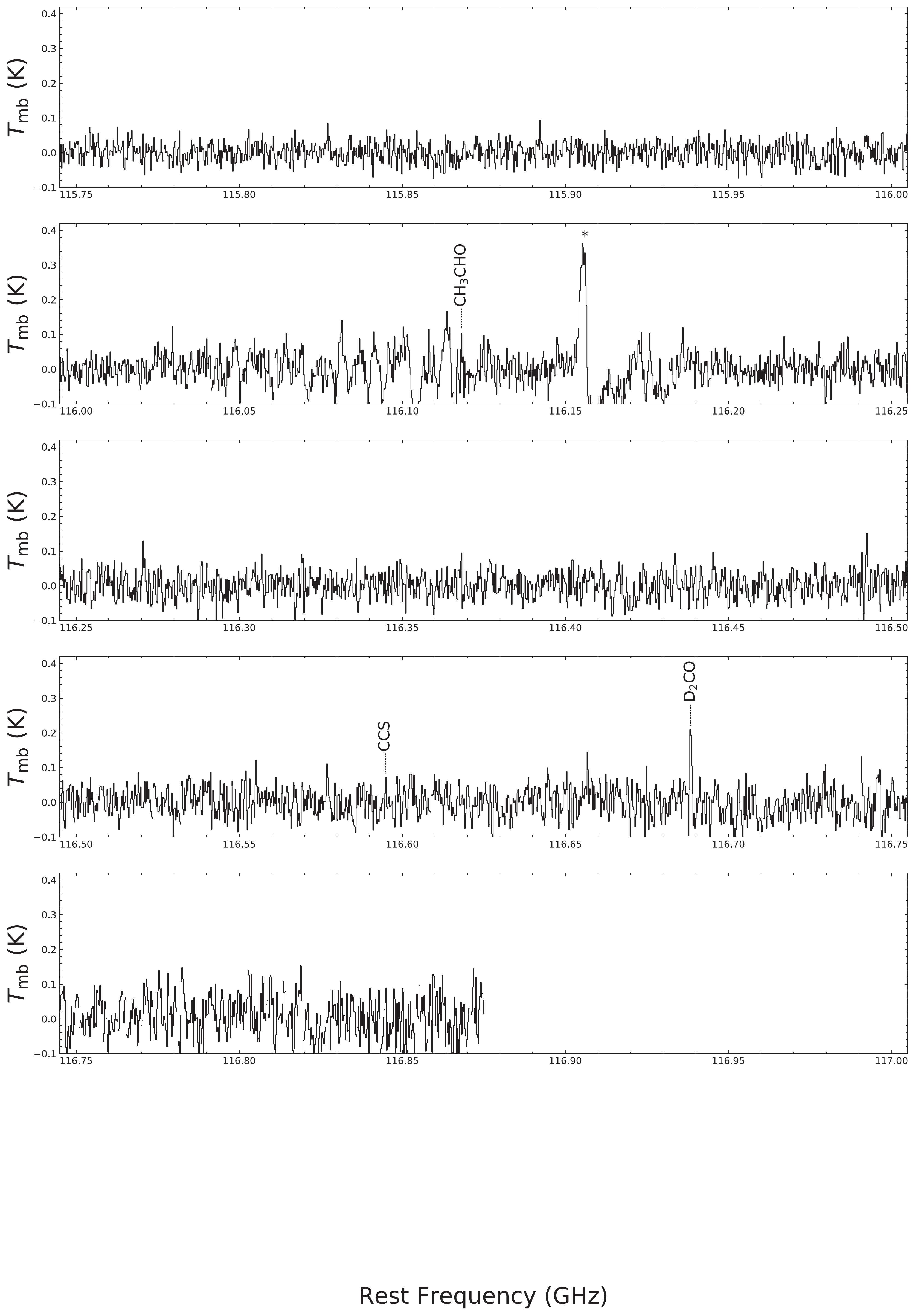}
\caption{(Continued.)}
\end{figure}
\clearpage
\begin{figure}[t]
\centering
 \includegraphics[height=.99\textheight]{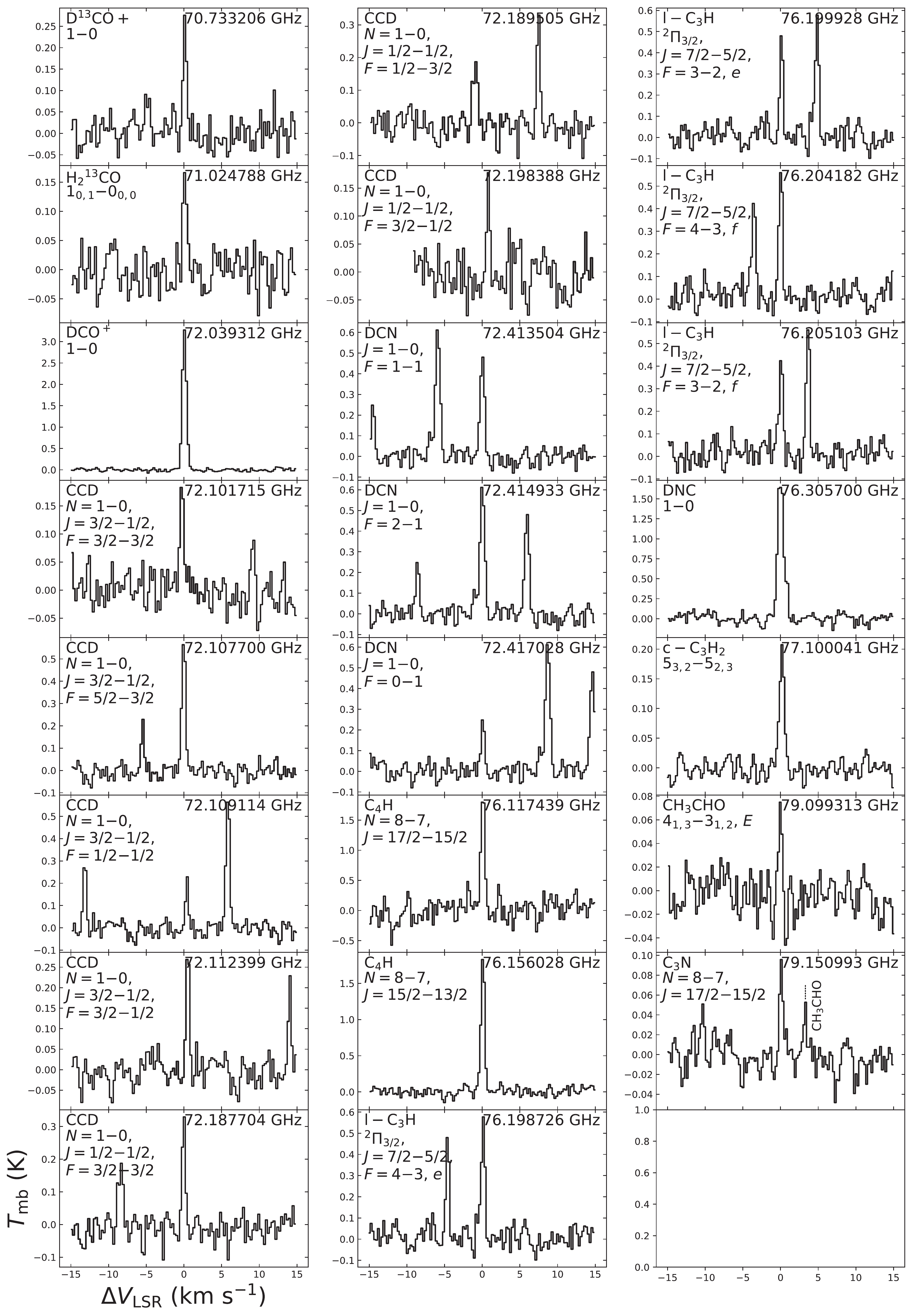}
\caption{Individual spectral line profiles of molecules detected in the 70 GHz band. The  $N=1$--$0$ lines of N$_2$D$^+$ are shown in Figure \ref{N2DCH2CN}.}\label{spect_4mm}
\end{figure}
\clearpage
\begin{figure}[t]
\centering
 \includegraphics[bb=0 0 1638 3495, height=\textheight]{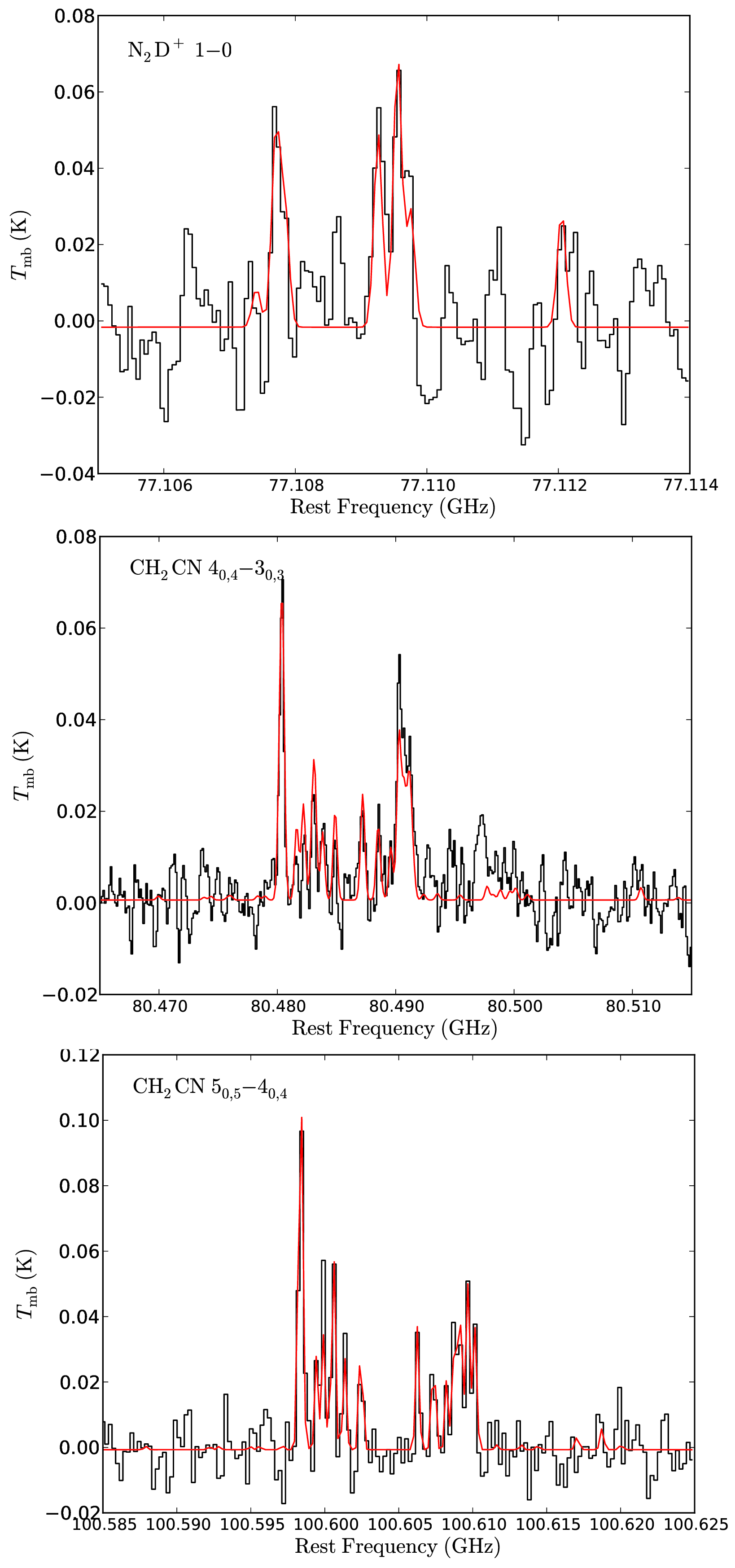}
\caption{The hyperfine components of N$_2$D$^+$ and CH$_2$CN. The results of the multiple Gaussian fitting are also shown in red.}\label{N2DCH2CN}
\end{figure}
\begin{figure}[t]
\centering
 \includegraphics[bb=0 0 4825 4658, width=.8\textwidth]{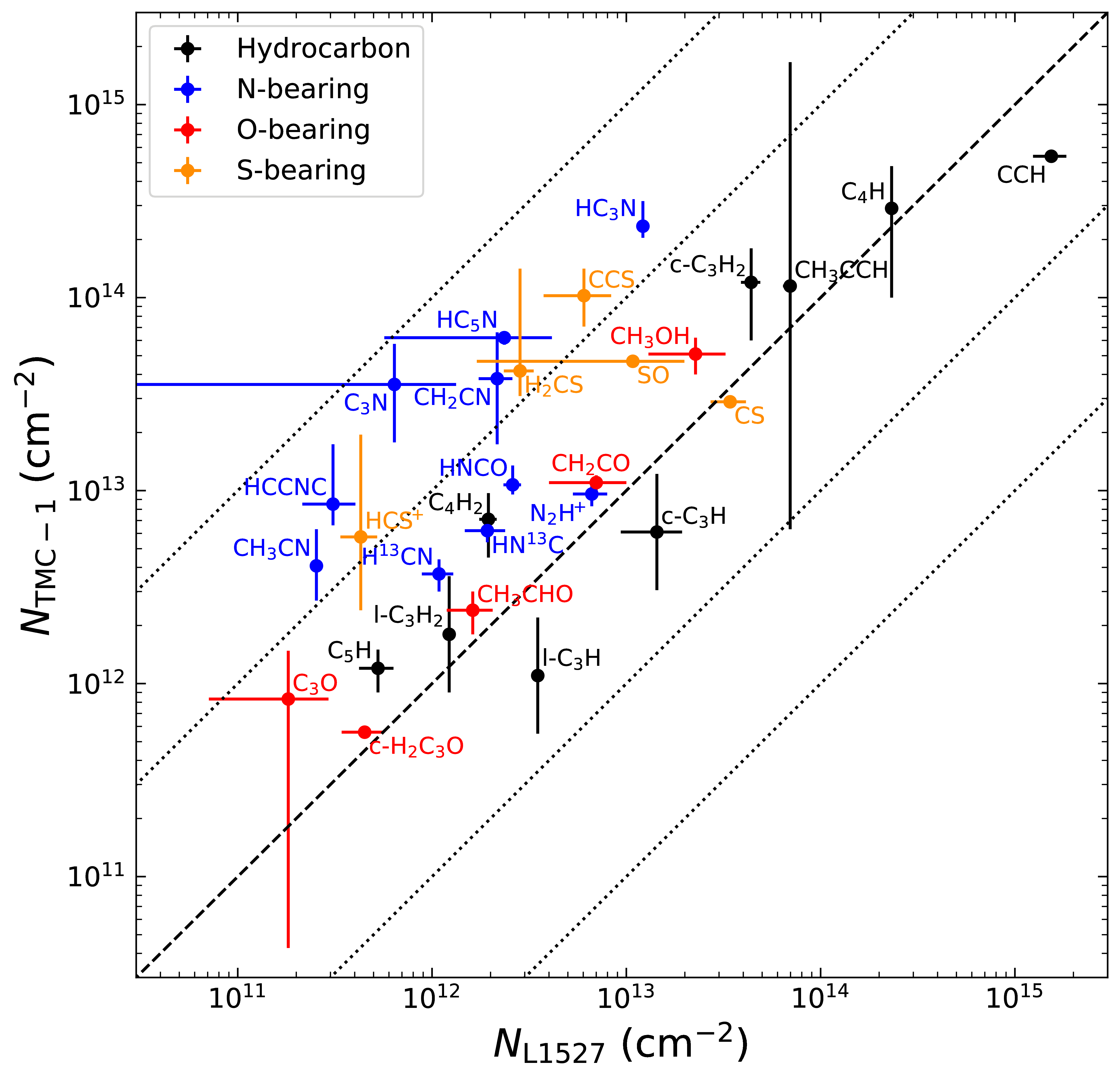}
\vspace{80pt}
\caption{The correlation plot of the column densities in L1527 ($N_\mathrm{L1527}$) and in TMC-1 ($N_\mathrm{TMC\mathchar`-1}$). The dashed line represents $N_\mathrm{L1527}=N_\mathrm{TMC\mathchar`-1}$. The four dotted lines show the column density ratio of 100, 10, 0.1, and 0.01. Column densities in TMC-1 are taken from:  Loison et al. (\yearcite{loison17}; c-C$_3$H, l-C$_3$H, c-C$_3$H$_2$, and l-C$_3$H$_2$), Sakai (\yearcite{s_phd}; C$_4$H, C$_4$H$_2$, and C$_5$H), Sakai et al. (\yearcite{s08b}; C$_4$H, C$_4$H$_2$, and C$_5$H, \yearcite{s10b}; CCH), Gratier et al. (\yearcite{gratier16}; CH$_3$CCH, HC$_3$N, HCCNC, CH$_2$CN, CH$_3$CN, HNCO, C$_3$N, CH$_2$CN, C$_3$O, SO, CS, CCS, H$_2$CS, and HCS$^+$), Taniguchi et al. (\yearcite{taniguchi16a}; HC$_5$N),  Soma et al. (\yearcite{soma15}; CH$_3$OH, \yearcite{soma18}; CH$_2$CO and c-H$_2$C$_3$O), Crapsi et al. (\yearcite{crapsi05}; N$_2$H$^+$), and Hirota et al. (\yearcite{hirota98}; HN\ss{13}C and H\ss{13}CN) }\label{N_tmc1}
\end{figure}
\clearpage
\begin{figure}[t]
\centering
 \includegraphics[bb=0 0 4825 4658, width=.8\textwidth]{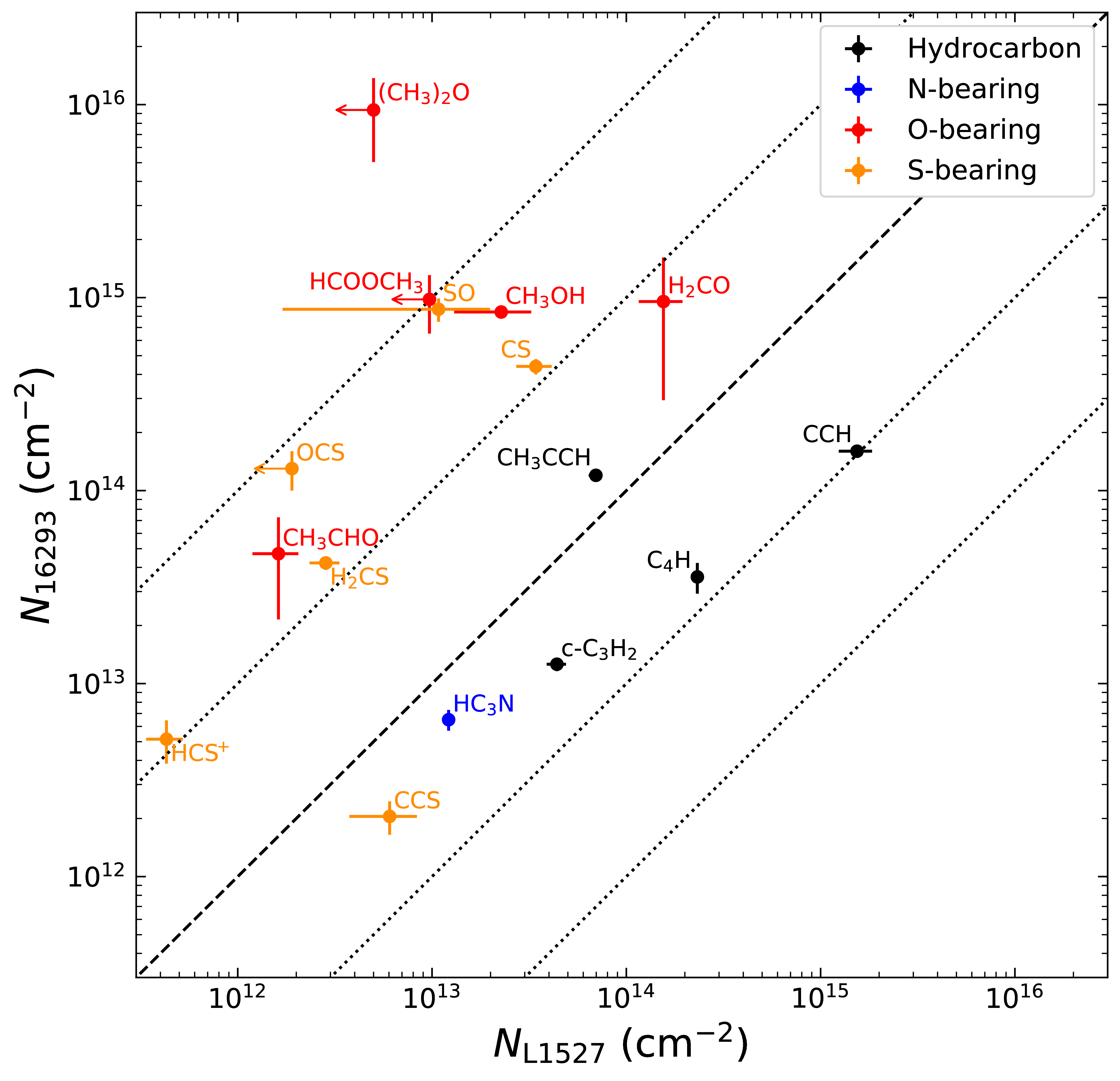}
\vspace{80pt}
\caption{The correlation plot of the column densities of carbon-chain molecules, COMs, and S-bearing molecules in L1527 ($N_\mathrm{L1527}$) and in IRAS 16293-2422 ($N_\mathrm{16293}$). The dashed line represents $N_\mathrm{L1527}=N_\mathrm{16293}$. The four dotted lines show the column density ratio of 100, 10, 0.1, and 0.01. The beam-averaged column densities in IRAS 16293-2422 are derived from \citet{cazaux03} for CH$_3$CHO, HCOOCH$_3$, and (CH$_3$)$_2$O. The column densities of other species are derived from \citet{caux11}.}\label{N_16293}
\end{figure}
\end{document}